\documentclass{aa}
\bibpunct{(}{)}{;}{a}{}{,} 
\usepackage{natbib}
\usepackage{graphicx}
\usepackage{epstopdf}
\usepackage{txfonts}
\usepackage{bm}
\usepackage{amsmath,amssymb}
\usepackage{booktabs} 

\usepackage[svgnames]{xcolor}
\definecolor{ku}{RGB}{144,26,30}
\definecolor{ku-yellow}{RGB}{255,249,25}
\definecolor{olive}{RGB}{107,142,35}
\definecolor{olive-yellow}{RGB}{154,205,50}

\begin{document}

\title{Water worlds in N-body simulations with fragmentation \\ in systems without gaseous giants}
   \author{A. Dugaro \inst{1,2}
           \thanks{adugaro@fcaglp.unlp.edu.ar},
           G. C. de El\'ia \inst{1,2}, \and
           L. A. Darriba \inst{1,2}}

   \offprints{A. Dugaro
    }

  \institute{Instituto de Astrof\'{\i}sica de La Plata, CCT La Plata-CONICET-UNLP \\
   Paseo del Bosque S/N (1900), La Plata, Argentina
   \and Facultad de Ciencias Astron\'omicas y Geof\'\i sicas, Universidad Nacional de La Plata \\
   Paseo del Bosque S/N (1900), La Plata, Argentina}
   
   \date{Received / Accepted}

  \abstract
   {}
{We analyze the formation and evolution of terrestrial-like planets around solar-type stars in the absence of gaseous giants. In particular, we focus on the physical and dynamical properties of those that survive in the system's Habitable Zone (HZ). This investigation is based on a comparative study between N-body simulations that include fragmentation and others that consider all collisions as perfect mergers.}
{We use an N-body code, presented in a previous paper, that allows planetary fragmentation. We carry out three sets of 24 simulations for 400 Myr. Two sets are developed adopting a model that includes hit-and-run collisions and planetary fragmentation, each one with different values of the individual minimum mass allowed for the fragments. For the third set, we considered that all collisions lead to perfect mergers.}
{The planetary systems produced in N-body simulations with and without fragmentation are broadly similar, with some differences. In simulations with fragmentation, the formed planets have lower masses since part it is distributed amongst collisional fragments. Additionally, those planets presented lower eccentricities, presumably due to dynamical friction with the generated fragments. Perfect mergers and hit-and-run collisions are the most common outcome.
Regardless of the collisional treatment adopted, most of the planets that survive in the HZ start the simulation beyond the snow line, having very high final water contents. The fragments' contribution to their mass and water content is negligible. Finally, the individual minimum mass for fragments may play an important role in the planets' collisional history.}
{Collisional models that incorporate fragmentation and hit-and-run collisions lead to a more detailed description of the physical properties of the terrestrial-like planets formed. We conclude that planetary fragmentation is not a barrier to the formation of water worlds in the HZ. }


\authorrunning{A. Dugaro,
           G. C. de El\'ia \and
           L. A. Darriba
               }
\titlerunning{N-body simulations with fragmentation without gaseous giants}

\maketitle
\section{Introduction}
\label{sec:introduction}

During the formation stage of a planetary system, the star and the protoplanetary disk can go through different stages. For instance, for the Solar System, it is believed that planetesimals formed in large numbers due to dust coagulation \citep{Krijt2015}. Alternatively, planetesimals could have formed by the gravitational collapse of concentrated, small particles \citep{Johansen2009}.

The growth beyond planetesimals occurs due to the enhancement of the gravitational focusing as the masses of the bodies become larger due to direct collisions between planetesimals. Pebble accretion contributes to this growth depending upon the abundance and size of surviving small solids. 
The planetesimals formation may lead to a runaway growth stage, in which larger objects grow considerably faster than smaller objects \citep{Kokubo1996}. This runaway growth slows down when the largest and most massive objects dominate the velocity evolution of smaller planetesimals. 

During the next stage, called oligarchic growth, each region of the disk is dominated by a single planetary embryo. Those embryos swept up nearby planetesimals \citep{Kokubo1998}. This stage ends when embryos approach the isolation mass and gravitational interactions between them become stronger. As a consequence of this, the embryos started to experiment orbits crossing and begin to scatter one another.

The final stage of terrestrial planet formation consists of gravitational interactions between the embryos and a remaining population of small planetesimals, where a late giant impact phase ends up forming the final planets. The occurrence of this last stage can play a key role since it shapes the final dynamical and physical properties, as well as the composition of the resulting planets.

It is worth mentioning that planetary systems that formed super-Earths could go through a different evolutionary path \citep{Lambrechts2019}. These objects, which have masses between those of the Earth and Neptune, experienced a different and more rapid growth process due to pebble accretion during the gaseous phase. After the disc dissipation, the system could undergo a gravitational instability and, subsequently, embryos could suffer a series of giant impacts reaching the final configuration of the system. For more detailed information on how pebble accretion and the effect of migration could play a key role in the formation of super-Earths see, for instance, \citet{Bitsch2019}, \citet{Izidoro2019}, and \citet{Lambrechts2019}.

The different physical and dynamical processes described above will yield a broad spectrum of planetary architectures. In this sense, several observational works \citep[e.g.][]{Cumming2008,Howard2013} and theoretical studies \citep[e.g.][]{Mordasini2009,Ida2013,Ronco2017} show a wide diversity of planetary systems in the Universe, and suggest that those only composed by terrestrial-like planets would seem to be the most common.

The study of terrestrial planet formation and their potential habitability has gained increasing relevance over the last few decades.
Because of this, it is necessary to develop a detailed model of the processes involved in the formation of terrestrial-like planets.
The late stages of terrestrial planet formation have been broadly studied using N-body simulations. These tools are appropriate to compute the evolution of this type of system. With new technologies and the advancement in the exoplanets' observations, the models of planetesimal accretion were applied to N-body simulations to study the formation of the new discoveries of exoplanetary systems. 

For the classic model of accretion using N-body numerical simulations, it is possible to study the dynamical evolution of planetary systems and to determine, in detail, the physical and orbital parameters of the planets. 
The physical properties of the bodies involved in the simulation are of utmost importance to compute the collisions between them.
For the sake of numerical simplicity and computational effort, many classic N-body integrators, such as the MERCURY package \citep{Chambers1999}, consider the collisions between planetary bodies as perfect mergers. This consideration implies that the two bodies involved in the collision blend into a single surviving object, with a mass equal to the sum of their masses, and preserving the total momentum. This approximation is acceptable when low-velocity collisions occur. On the contrary, when the collisions take place at high velocities (namely, a few times the mutual escape velocity of the involved bodies), this consideration is no longer valid, since mass loss is more likely to happen. 

Despite the limitations concerning perfect mergers used by classical integrators, this model of accretion successfully reproduces the general aspects of the terrestrial planets of our solar system and their overall dynamics \citep{Chambers2001,Raymond2006,Obrien2006,Raymond2009}.
Moreover, for extrasolar systems, different works also used the classical accretion model to study terrestrial planet formation around stars of different spectral types, both with or without gas giants \citep[e.g.][among others]{deElia2013,Dugaro2016,Darriba2017,Sanchez2018,Zain2018}.

To have a more realistic computation of planetary formation, we have to take into consideration the possibility of bodies' fragmentation during a collision. 
However, modeling this process with an N-body code has several aspects to be considered. 
One of the most important factors is the number of fragments generated in each collision. Following \citet{Chambers2013}, this quantity will depend on a given parameter related to the minimum mass allowed for the fragments to have ($M_\text{min}$). This value must be chosen in such a way that sets a compromise between the computing time and the realism that we want for the simulations. On the one hand, a low value of $M_\text{min}$ will produce an excessive number of bodies, making the simulations unfeasible, given that collisional fragmentation can increase the number of objects, slowing down the time required to complete a simulation. On the other hand, considering a value of $M_\text{min}$ too large will overestimate partial and perfect accretions. It is worth mentioning that, due to the stochastic nature of N-body simulations with $N > 2$, a large number of numerical experiments have to be carried out to get successful statistical results.

Recently, \citet{Leinhardt2012} developed a complete prescription for more realistic collisions that includes fragmentation for gravity-dominated bodies. This prescription describes that the possible outcomes for a given collision depend on geometric and physical parameters, such as the target's and projectile's mass, impact angle, and collisional velocity. The number of fragments that could be generated in a given collision depends on the specific impact energy per unit mass of the system ($Q$) and a threshold value ($Q^*$), defined as the specific energy needed to scatter half of the total colliding mass. It is worth mentioning that authors such as \citet{Cambioni2019} and \citet{Gabriel2020} presented several improvements and updates regarding the parameters space mapping. The transition between hit-and-run collisions and merging criteria for collisions between rocky planets was studied by \citet{Genda2012}. These authors found that, if a hit-and-run collision has an impact velocity lower than a given critical velocity, a second collision occurs, leading to a perfect merge and, thus, re-accreting the material.

Recently, several authors moved away from the standard accretion model and used a refined treatment for the collisions. \citet{Chambers2013} implemented the fragmentation model into the well-known MERCURY package \citep{Chambers1999} and studied terrestrial planets formation, comparing a set of simulations with the perfect accretion model to another with the refined treatment of the collisions. This improvement was included in an N-body tree code, as in the works developed by \citet{Lines2014} and \citet{Bonsor2015}, who analyzed planetesimal formation. Moreover, \citet{Quintana2016} used the modified version of MERCURY that allows fragmentation to study how giant impacts affect terrestrial planet formation. For that matter, the authors compared two sets of initial conditions and performed simulations with and without fragmentation. The authors concluded that the final architecture for the systems are comparable, while the collisional history of the resulting planets differs significantly. Then, \citet{Wallace2017} performed a series of N-body simulations and analytical investigations of rocky planet formation at small semimajor axes, where tidal effects become important. They concluded that collisional fragmentation is not a barrier to terrestrial planet formation, except at distances within 10\% of the Roche radii. Moreover, the authors performed different sets of N-body simulations, varying the minimum permitted mass for the fragments, $M_{\text{min}}$. From this, \citet{Wallace2017} found out that the final number of planets and their masses do not significantly change with such a parameter. In this line of investigation, \citet{Childs2018} performed N-body simulations with fragmentation and studied how the variation of the gaseous giant's mass affects the final architecture of a planetary system, its formation timescale, and the collisional history of the planets formed. The authors found out that the inclusion of fragmentation has a minor impact on these features. Additionally, they found no correlation with the giant planets' masses and the number of Earth analogs produced in their simulations. \citet{Mustill2018} studied in-situ planet formation on packed systems and instability scenarios with their implementation of the collisional improvement developed by \citet{Leinhardt2012}. Additionally, the authors introduced a mass removal factor to represent material grounded and then removed from the system by radiation forces. For the solar system, \citet{Clement2019} presented a set of simulations of terrestrial planet formation using an integrator that considers the effects of collisional fragmentation. They concluded that compared with simulations without fragmentation, the systems that include a refined treatment of collisions provide better matches to the Solar System’s terrestrial planets in terms of their dynamical evolution and compact mass distribution. Collisional processes also tend to lengthen the dynamical accretion timescales of Earth analogs and shorten those of Mars analogs. Additionally, the authors found that collisional fragmentation and hit-and-run collisions play a dominant role in preventing planet formation in the primordial asteroid belt. 
\citet{Kobayashi2019}, investigated the orbital evolution of protoplanets in a planetesimal disk taking into account collisional fragmentation of planetesimals. Due to collisional cascade, the planetesimals are ground to dust and blown away by radiation pressure. This process weakens dynamical friction. Because of this phenomenon, for small planetesimals, the dynamical friction is insignificant for the planets' eccentricity damping due to collisional fragmentation.
In addition, they studied the orbital evolution of individual planets in a swarm of ejecta produced by giant impacts. The authors found that giant impact ejecta plays a primary role in determining the orbits of terrestrial planets. \citet{Deienno2019} tested different values for the coefficient of restitution for collisions and studied the energy dissipation within embryo-embryo impacts. Their results show that varying the dissipated energy level within embryo–embryo collisions do not play an important role in the final terrestrial planetary system. 

Recently, in \citet{Dugaro2019}, we presented an N-body code called \emph{D3} that allows planetary fragmentation for gravity-domain bodies based on the works of \citet{Chambers1999}, \citet{Leinhardt2012}, \citet{Genda2012}, \citet{Chambers2013}, and \citet{Mustill2018}. The authors studied terrestrial planet formation on a protoplanetary disk, hosted by a solar-type star, and the presence of Jupiter and Saturn analogs. The authors compared two sets of identical initial conditions and performed a series of N-body simulations with and without fragmentation. The results derived in \citet{Dugaro2019} are consistent with those obtained by \citet{Chambers2013} and \citet{Genda2012}. 

The use of N-body simulations that include fragmentation allows us to perform a more detailed study of the final composition of the planets formed. In particular, we can study the water loss/accretion of the final planets more realistically than in the classic models of accretion.
\citet{Marcus2010} presented two empirical models for the mantle stripping in differentiated planetary embryos after a collision. The authors set a simple planet structure of two layers, assuming differentiation in core/mantle, where the mantle could be composed by silicate or ice. In this work, the authors concluded that the more energetic the collision, the more mass from the mantle is lost. Therefore, for violent collisions, water could be more easily removed. \citet{Dvorak2015} performed SPH (Smoothed Particle Hydrodynamics) simulations and studied water loss in planetary embryos and water retained in significant fragments after a collision. They concluded that the impact velocity and the impact angle play a key role in the water loss of a planetary embryo after a collision. The investigations developed by \citet{Marcus2010} and \citet{Dvorak2015} suggest that incorporating a realistic model of volatile transport and removal in an N-body code, may lead to reduced water contents on the resulting terrestrial-like planets, in comparison with those derived from classical models that assume perfect mergers.
\citet{Burger2018} studied the volatile loss and transfer. The authors focused on hit-and-run encounters using SPH simulations. They concluded that the cumulative effect of several hit-and-run collisions could efficiently strip off volatile layers of protoplanets.
Driven by this, \citet{Dugaro2019}  studied the water delivery in planets formed in the habitable zone (HZ), using the mantle stripping models derived by \citet{Marcus2010} in their N-body simulations with fragmentation.
The authors showed that fragmentation is not a barrier for the surviving of water worlds in the HZ, and fragments may be important in the final water content of the potentially habitable terrestrial planets formed in situ. 

The main goal of the present research is to analyze the physical and dynamical properties of terrestrial-like planets and water delivery in the HZ, in the absence of gaseous planets. This analysis is carried out during the late-stage accretion phase of terrestrial planet formation. In addition to this, we study the survival of water worlds through N-body simulations that incorporate fragmentation and hit-and-run collisions. Moreover, we study the sensitivity of the overall result to the minimum mass allowed for the fragments $M_\text{min}$.

This paper is organized as follows. Section \ref{sec:num_model} contains a basic description of the numerical model used in this work. Section \ref{section:applications} describes the initial conditions selected, as well as the properties of the protoplanetary disk used in this paper. The main results are presented in Sect. \ref{section:results}. We address the model's limitations in Sect. \ref{sec:limitations}. Lastly, Sect. \ref{section:discusion} contains the main conclusions  of the present research.


\section{Numerical model}
\label{sec:num_model}

In a previous paper \citep{Dugaro2019}, we described in detail the construction of an N-body code, which included a refined collisional treatment for the planetary embryos. In this section, we briefly present the highlights of the collisional model implemented and show the basic ideas of the analysis made whenever a collision between two planetary embryos takes place.

To determine the outcome of a collision between two massive bodies, we need to calculate the impact energy per unit mass $Q$, given by
  \begin{equation}
    Q = \frac{1}{2}\mu\frac{v_\text{i}^2}{M_\text{t}+m_\text{p}},
  \label{Eq1}    
  \end{equation}
\noindent{where} $M_\text{t}$ is the mass of the target, $m_\text{p}$ the mass of the projectile, $v_\text{i}$ the impact velocity, and $\mu$ the reduced mass, which is calculated as follows:
 \begin{equation}
    \mu = \frac{ M_\text{t}m_\text{p}}{M_\text{t}+ m_\text{p}}.
  \end{equation}

Following \citet{Leinhardt2012}, we need to compare $Q$ given by Eq. \eqref{Eq1} with a threshold value $Q^*$, which is defined as the specific impact energy needed to disrupt half of the total colliding mass. Depending on this comparison, \citet{Leinhardt2012} derived analytic expressions to determine the different possible outcomes for the collisions as well as different expressions for the mass of the largest remnant ($M_\text{lr}$) after the collision.

The regimes derived by \cite{Leinhardt2012} and used in \citet{Dugaro2019} are:
\begin{itemize}
\item\textbf{Perfect merging}, where the mass of the largest remnant $M_\text{lr}$ is the sum of the mass of the target $M_\text{t}$ and that of the projectile $m_\text{p}$,
\item\textbf{Partial accretion}, where $M_\text{lr}$ $<$ ($M_\text{t}$ + $m_\text{p}$) and $M_\text{lr}$ > $M_\text{t}$, 
\item\textbf{Erosive collision}, where $M_\text{lr}$ $<$ ($M_\text{t}$ + $m_\text{p}$) and  $M_\text{lr}$ $<$ $M_\text{t}$, 
\item\textbf{Super-catastrophic collision}, where $M_\text{lr}$ $<$ 0.1$M_\text{t}$,
\item\textbf{Hit-and-run}, where $M_\text{t}$ and $m_\text{p}$ remain unaltered,
\item\textbf{Graze and merge}, where a hit-and-run occurs with a second collision that leads to a perfect merging,
\item\textbf{Erosive hit-and-run}, where $M_\text{t}$ remains unaltered but the projectile may suffer fragmentation.
\end{itemize}

In those collisions that do not lead to a perfect merge or a hit-and-run, the mass of the system of colliding bodies is distributed in a largest remnant of mass $M_\text{lr}$ and in the several fragments produced. Following \cite{Chambers2013} and \citet{Dugaro2019}, we distribute the total mass of the fragments $m_\text{frag}$ in equally sized bodies. According to this, $m_\text{frag}$ = $M_\text{t}$ + $m_\text{p}$ - $M_\text{lr}$, and the number of fragments is calculated as $m_\text{frag}$/$M_\text{min}$, being $M_\text{min}$ the minimum permitted individual mass for the fragments. One of the scopes of this paper is to analyze how the modification of this parameter may affect the overall dynamics and physical properties of the planets formed in our N-body experiments. Since reducing the value of $M_\text{min}$ will increase the number of fragments generated in an impact event, the selection of this value sets a compromise between the number of fragments simulated in the model and the computing performance \citep{Chambers2013,Dugaro2019}.
Following Chambers (2013), the fragments generated in a collision are assigned to have an absolute value for their velocity of 5\% above the mutual escape velocity of the combined target and projectile system. For the direction of the fragments' velocity, they are randomly generated in a way that guarantees the conservation of linear momentum. It is worth noting that \citet{Clement2019b} performed several N-body simulations that included collisional fragmentation. In their research, they varied the speed at which the fragments are ejected after a given collision. The authors concluded that increasing the ejection speed, leads to a lower probability of being re-accreted by the target body. Moreover, increasing the speed also increases the chances of fragments' loss due to collisions with the star or being ejected from the system itself.

For all the collisions that take place in a simulation, the \emph{D3} code records the initial and final positions and velocities of the bodies, together with geometric and physical parameters such as the impact angle and collisional velocity. All this information is used in a post-process stage that allows us to characterize the overall behavior of the system and the collisional evolution of the planets that survive in the system.
The reader can find a detailed discussion about the regimes of collision and their modeling in the N-body code in \citet{Dugaro2019}.


\section{Initial conditions - Setup}
\label{section:applications}

In this section, we present the numerical model and physical parameters that define the protoplanetary disk of our work. Moreover, we specify the initial conditions to be used in our N-body experiments.

We implemented the same initial disk model that was used by \citet{Dugaro2019}, using a gas-surface density profile $\Sigma_\text{g}(R)$ and a solid-surface density profile $\Sigma_{\text{s}}(R)$ given by \citep{Lynden1974,Hartmann1998}:
\begin{equation}
\Sigma_{\text{g}}(R) = \Sigma_{\text{g}}^{0}\left(\frac{R}{R_{\text{c}}}\right)^{-\gamma} \text{exp}\left[-\left(\frac{R}{R_{\text{c}}}\right)^{2-\gamma}\right],
\label{eq:gas}
\end{equation}
\begin{equation}
\Sigma_{\text{s}}(R) = \Sigma_{\text{s}}^{0}\eta_{\text{ice}}\left(\frac{R}{R_{\text{c}}}\right)^{-\gamma} \text{exp}\left[-\left(\frac{R}{R_{\text{c}}}\right)^{2-\gamma}\right],
\label{eq:solid}
\end{equation} 
where $R$ is the radial coordinate in the disk's mid-plane, $R_{\text{c}}$ a characteristic radius, $\Sigma_{\text{s}}^{0}$ and $\Sigma_{\text{g}}^{0}$ normalization constants, and $\gamma$ an exponent that determines the density gradient. The $\eta_\text{ice}$ parameter in Eq. \eqref{eq:solid} represents an increase in the amount of solid material due to the condensation of volatile material beyond the snow line. 

To determine the values of the parameters $\Sigma_{\text{s}}^{0}$ and $\Sigma_{\text{g}}^{0}$, we adopted the same considerations made by \citet{Dugaro2019}. The value of $\Sigma_{\text{g}}^{0}$ is calculated by integrating Eq. \eqref{eq:gas} over the total disk area assuming axial symmetry. From this, $\Sigma_{\text{g}}^{0}=(2-\gamma)M_{\text{d}}/(2 \pi R^{2}_{\textrm{c}})$, being $M_{\textrm{d}}$ the total mass of the disk. Then, for $\Sigma_{\text{s}}^{0}$, we compute it as $\Sigma_{\text{s}}^{0}$ = $z_\text{0}$ $\Sigma_{\text{g}}^{0}$ $10^{[\textrm{Fe/H}]}$, where [\textrm{Fe/H}] is the stellar metallicity and $z_\text{0}$ the primordial abundance of heavy elements in the Sun, which has a value of 0.0153 \citep{Lodders2009}. 

As in \citet{Dugaro2019}, we assumed that the central star has a mass of $M_{\star} = 1  \text{M}_{\odot}$ and solar metallicity. The values for $M_{\textrm{d}}$, $\gamma$, and $R_\textrm{c}$ adopted in this work were of 0.01 M$_{\odot}$, 0.9, and 25 au, respectively, which are consistent with several observational studies of protoplanetary disks in different star-forming regions \citep[e.g.][]{Andrews2010,Tazzari2017,Cieza2019}, and with analysis of disks based on population synthesis models \citep{Bate2018}. For the snow line, its determination is still a very active research field \citep[see, for instance,][]{Morbidelli2016}. For our work, following \citet{Ida2004}, we assume it is located at 2.7 au.\\

Our research will focus on the potential habitability of the resulting planets in our scenario of study. Thus, we must specify the limits of the HZ, which is defined as the region around the star in which a planet could retain liquid water on its surface. \citet{Kopparapu2013a} established inner and outer limits for the HZ around stars of different spectral types. In particular, the investigation to be developed in the present work made use of the so-called optimistic estimates for a solar-type star derived by those authors, where the inner (outer) edge is located at 0.75 (1.7) au. 

The development of the N-body experiments required the specification of physical and orbital parameters for the planetary embryos that initially composed the disk. For obtaining the initial conditions, we selected a set of planetary embryos following, as we did in \citet{Dugaro2019}, a semi-analytic study of oligarchic growth carried out by \citet{Kokubo1998, Kokubo2000}. We are aware that, to perform a more in-depth study of the processes involved in planetary formation, it is necessary to consider a more realistic set of initial conditions, like those obtained in works like \citet{Carter2015} and \citet{Walsh2019}. However, this is not the scope of this project, since we aim to analyze the effects of the implementation of a fragmentation regime in the formation of Earth-like planets. Because of this, we consider that, as it can be seen in other works in the literature \citep[see, for instance, ][]{Izidoro2017, Raymond2018}, among others), a simpler distribution like the one adopted in this work suffices.\\

\begin{figure}
 \centering
 \includegraphics[angle=0, width= 0.48\textwidth]{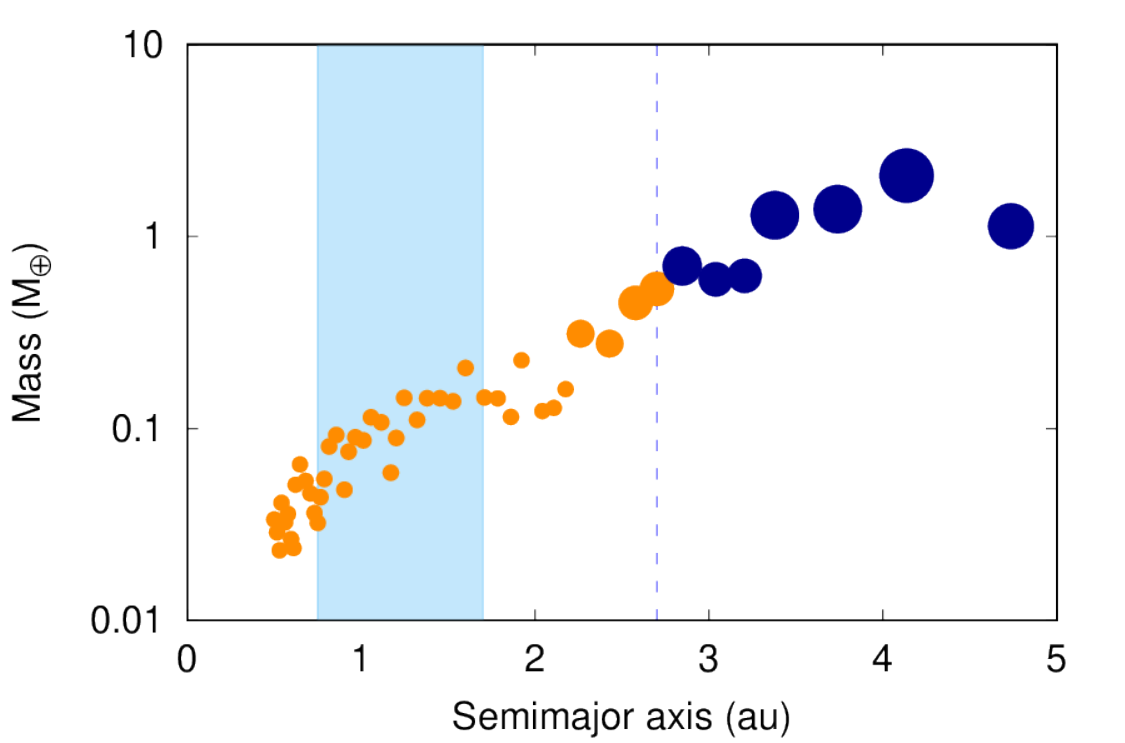}
 \caption{Mass distribution of planetary embryos from one representative simulation as a function of the initial semimajor axis at the end of the gaseous phase. The size of the circles is scaled with the mass of each embryo. The color code refers to the initial fraction of water by mass of the embryos. The orange (dark blue) circles indicate a fraction of 10$^{-4}$ (0.5) of water by mass. The sky-blue shaded region illustrates the HZ of the system, while the dashed blue line represents the snow line assumed in our model.}
\label{fig:distr_emb}
\end{figure}

As \citet{Dugaro2019}, the numerical simulations of the present work started with $\sim$50 embryos between 0.5 au and 5 au and masses ranging from 0.03 M$_{\oplus}$ to 1.13 M$_{\oplus}$. Such planetary embryos were distributed following the solid-surface density profile $\Sigma_{\text{s}}(R)$ given by Eq. \eqref{eq:solid}. The reader can find a detailed explanation of how to determine the mass and semimajor axis distributions for the embryos in \citet{Dugaro2019}. Because of the condensation of volatile material beyond the snow line, all bodies that begin the simulation in the region interior to 2.7 au were assumed to have a physical density of 3 g cm$^{-3}$, while those bodies initially located beyond 2.7 au were assigned to have a density of 1.5 g cm$^{-3}$. As for the orbital parameters, all planetary embryos were assumed to start the numerical simulations on quasi-circular and co-planar orbits. For each body, orbital eccentricities and inclinations were taken randomly from a uniform distribution with maximum values of 0.02 and 0.5$^\circ$, respectively. The remaining orbital elements such as the argument of the pericenter $\omega$, the ascending node longitude $\Omega$, and the mean anomaly $M$ were chosen randomly following a uniform distribution between 0$^{\circ}$ and 360$^{\circ}$. Figure \ref{fig:distr_emb} shows the initial mass distribution of the embryo population as a function of the initial semimajor axis. In the present research, we assumed that the physical and orbital parameters described above represent the initial conditions once the gas has been fully dissipated from the system.

The values proposed for the $\eta_{\text{ice}}$ parameter in Eq. \eqref{eq:solid} are associated with a radial compositional gradient in the protoplanetary disk due to the condensation of volatile material beyond the snow line. In particular, we considered that the water content by mass varies with the radial coordinate in the disk mid-plane $R$. As stated in \citet{Lodders2003}, in the solar system, the ice-to-rock ratio is approximately 1. This leads to an increase in the amount of solid mass by a factor of 2 with respect to the region interior to this line. Following this, we assume in Eq. \eqref{eq:solid} a value of $\eta_{\rm{ice}}=1$ for $R<R_{\rm{ice}}$ and $\eta_{\rm{ice}}=2$ for $R>R_{\rm{ice}}$. With this assumption, we adopted a water content fraction of 0.5 for all the embryos whose accretion seed location was initially beyond the snow line, while we assumed a value of $10
^{-4}$ for the water fraction of those interior to the snow line. Such a distribution of water was assigned to each planetary embryo of our N-body experiments based on its starting location. How the rocky planets located in the HZ acquired their final water content is a complex process. In fact, it may involve several transport mechanisms \citep[for a detailed description of these processes, see][]{Meech2019}. In this work, we considered that the water delivery to embryos in the region interior to the snow line was, mainly, through the migration of water-rich material formed beyond the aforementioned line.

An additional parameter that is necessary to perform simulations with fragmentation is the minimum individual mass $M_{\text{min}}$ to be assigned to the fragments. 
One of the scopes of this work is to analyze how the value of $M_\text{min}$ modifies the overall dynamics of a given system, as well as the physical properties of the terrestrial-like planets that could survive. To do this we carry out three different sets of N-body experiments as follows:

\begin{itemize}
\item Set F1: Simulations with fragmentation, using a value of $M_\text{min}$ = $M_\text{F1}$ = 0.018 M$_{\oplus}$,
\item Set F2: Simulations with fragmentation, using a value of $M_\text{min}$ = $M_\text{F2}$ = 0.0018 M$_{\oplus}$
\item Set NoF: Simulations without fragmentation, using the standard accretion model.
\end{itemize}

Because of the stochastic nature of the accretion process, we performed 24 N-body simulations for each set. The total integration time for all N-body experiments was 400 Myr \citep{Lykawka2019}.

The main results derived from our investigation are presented in the next section. 


\section{N-body simulations: Results}
\label{section:results}

In this section, we present the results of our 72 N-body experiments, which are associated with the sets of simulations F1, F2, and NoF. 

\subsection{General results}

\begin{figure}[ht!]
\centering
\includegraphics[angle=0, width= 0.48\textwidth]{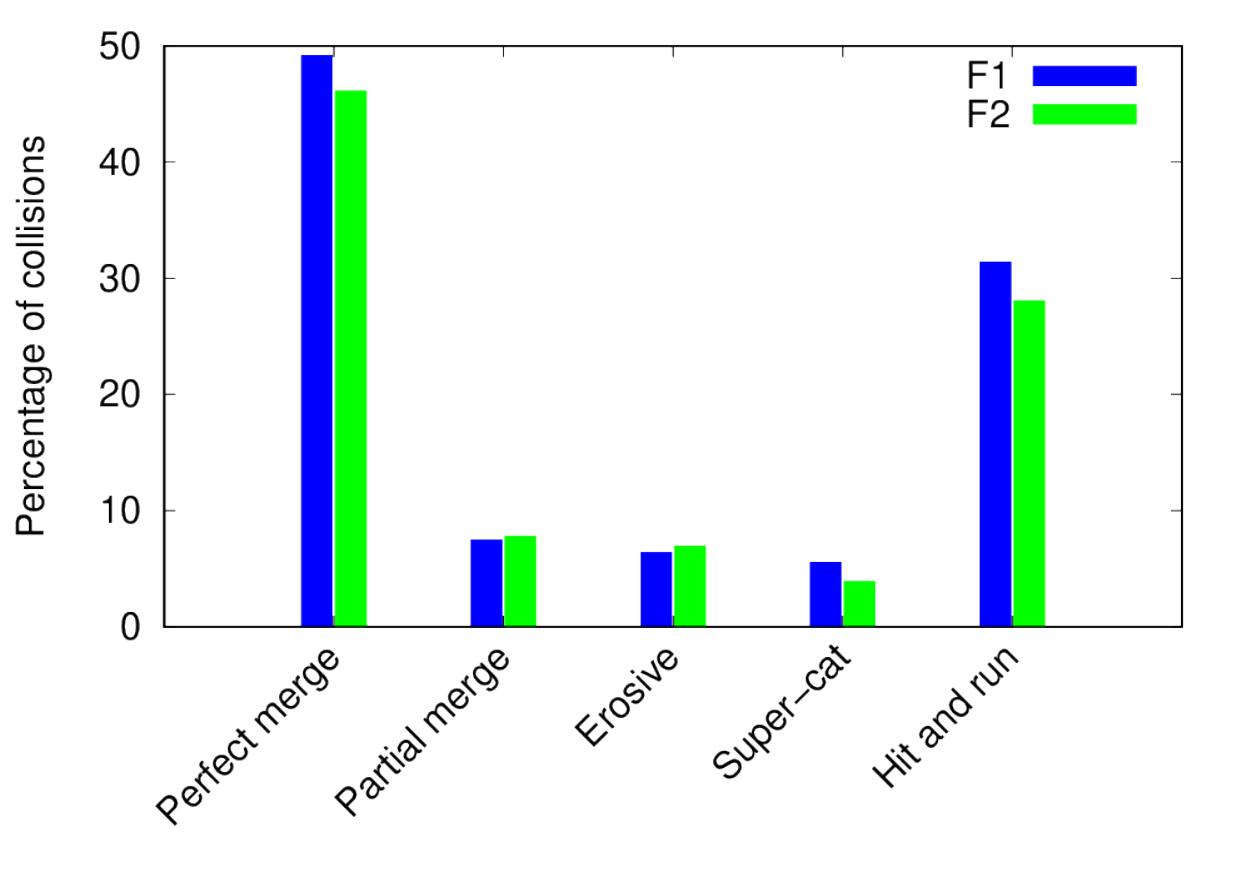}
\caption{Comparative histogram for the different collision regimes and their respective percentages of occurrence for the F1 (blue bars) and F2 (green bars) scenarios. It is important to remark that the collisions considered to build these histograms included only giant impacts. In fact, the collisions between embryos and fragments are not accounted for.} 
\label{fig:histograma}
\end{figure}

As a starting point, we analyzed the frequency of occurrence of the different collision types presented in Sect. \ref{sec:num_model}, in those N-body simulations that included fragmentation. Fig. \ref{fig:histograma} shows the percentages of the different collision types that planetary embryos experienced in sets F1 and F2, which are represented by blue and green boxes, respectively. This figure allows us to observe two important results. First, the collision frequency does not seem to be sensitive to the minimum individual mass of the fragments $M_{\text{min}}$, since there are no substantial differences concerning the percentage of collision types between sets F1 and F2. Second, the collisions show a bi-modal distribution in both scenarios, where most of them are clumped in perfect mergers ($\sim$48\%) and hit-and-run ($\sim$30\%). It is worth remarking that this last percentage also takes into account erosive hit-and-run collisions.

The percentages of hit-and-run collisions derived from sets F1 and F2 were significantly lower than those presented in \citet{Genda2012}, \citet{Chambers2013}, and \citet{Dugaro2019}, who obtained a value of $\sim$45\%. We consider that this difference could be attributed to the presence of giant planets in those three different studies, which favored the dynamical excitation of the planetary embryos, increasing the impact angle. 
The remaining $\sim$20\% of the collisions experienced by the embryos in sets F1 and F2 were partial mergers, erosive impacts, and super-catastrophic collisions. These types of impacts could lead to a substantial increase in the number of bodies, depending on the value of $M_{\text{min}}$. For that matter, it is interesting to study the number of bodies as a function of the integration time.

\begin{figure}[ht!]
 \centering
 \includegraphics[angle=0, width= 0.48\textwidth]{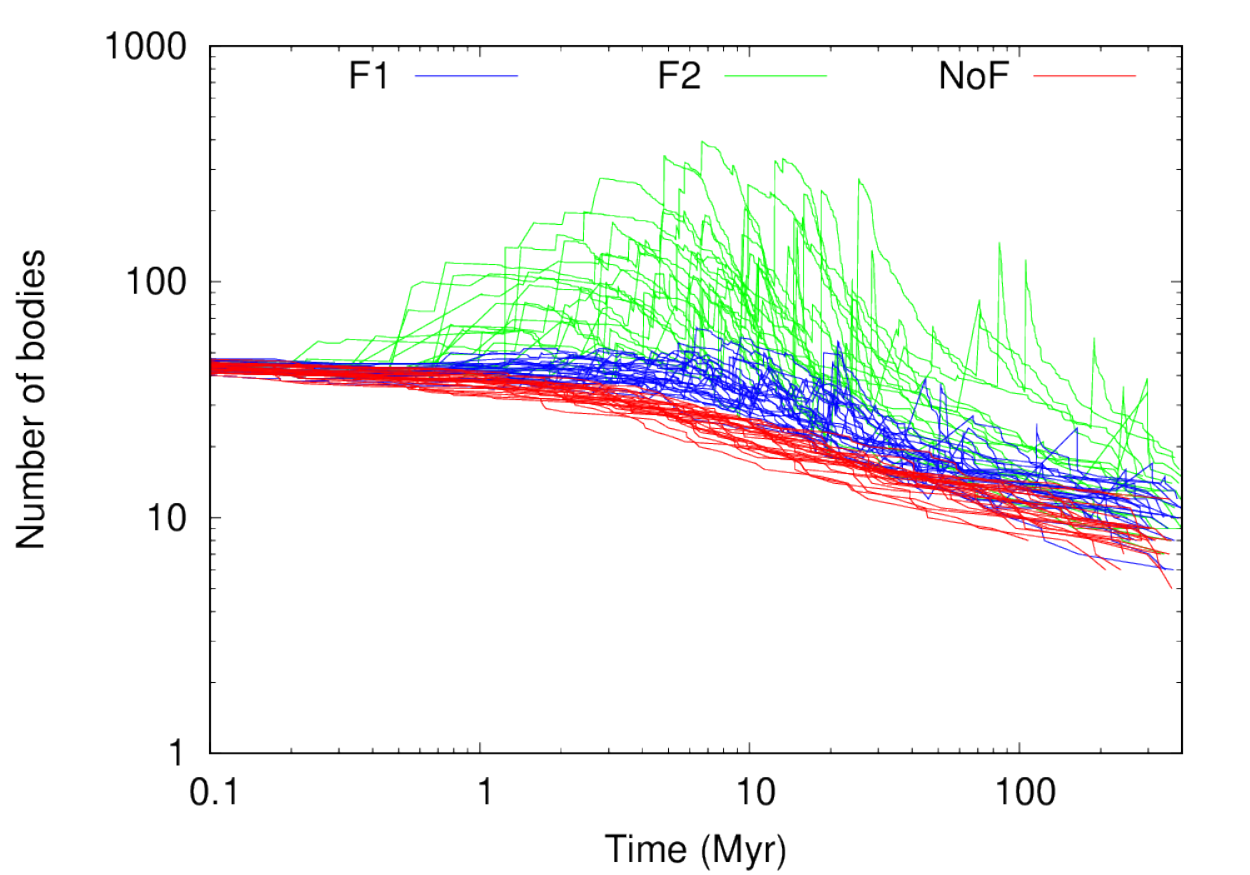} \\
 \includegraphics[angle=0, width= 0.48\textwidth]{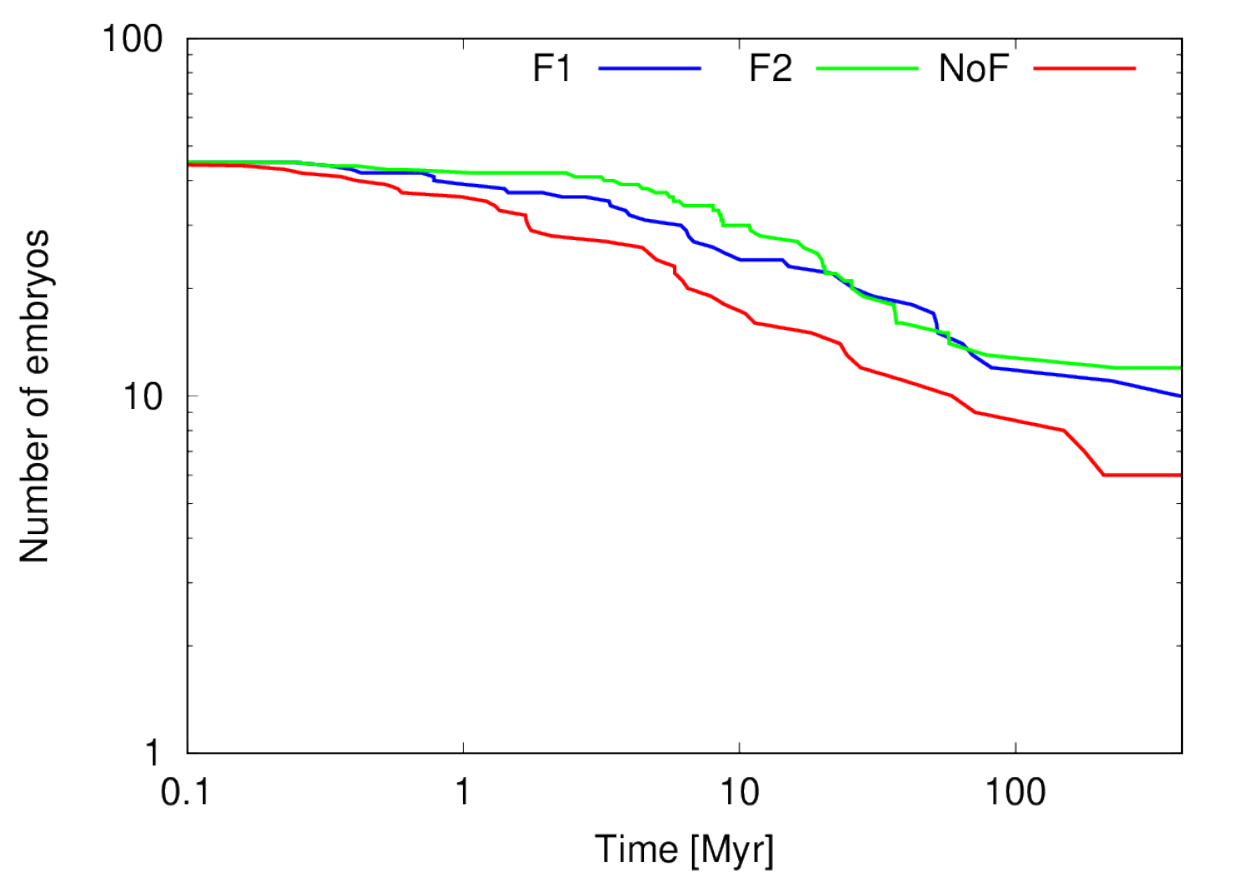}
 \caption{Time evolution for the number of bodies in sets F1, F2 and NoF, which are depicted with blue, green, and red traces, respectively. The top panel was generated considering both embryos and fragments in the system for all simulations. The bottom panel is analogous to the top panel, but only counting the number of embryos for a representative simulation.}
\label{fig:nvst_ST}
\end{figure}

The top panel of Fig. \ref{fig:nvst_ST} shows the temporal evolution of the total number of bodies in the numerical simulations of sets F1, F2, and NoF, which are illustrated as blue, green, and red curves, respectively. As expected, the curves associated with NoF simulations exhibit a monotonically downward trend, since all collisions were treated as perfect mergers. For this set, the total number of bodies in the numerical experiments reached half of its initial value in about 10 Myr. The situation is more complex for F1 and F2 scenarios. The blue curves associated with set F1 also show a downward trend, which is interrupted when fragment-generating collisions occur. These events produced instantaneous bumps in the number of bodies. These instantaneous increments turned out to be more significant in F2 simulations. The lower the value of $M_\text{min}$, the greater the number of fragments generated in the different impact events.

\begin{figure*}[ht!]
 \centering
 \includegraphics[angle=0, width= 0.33\textwidth]{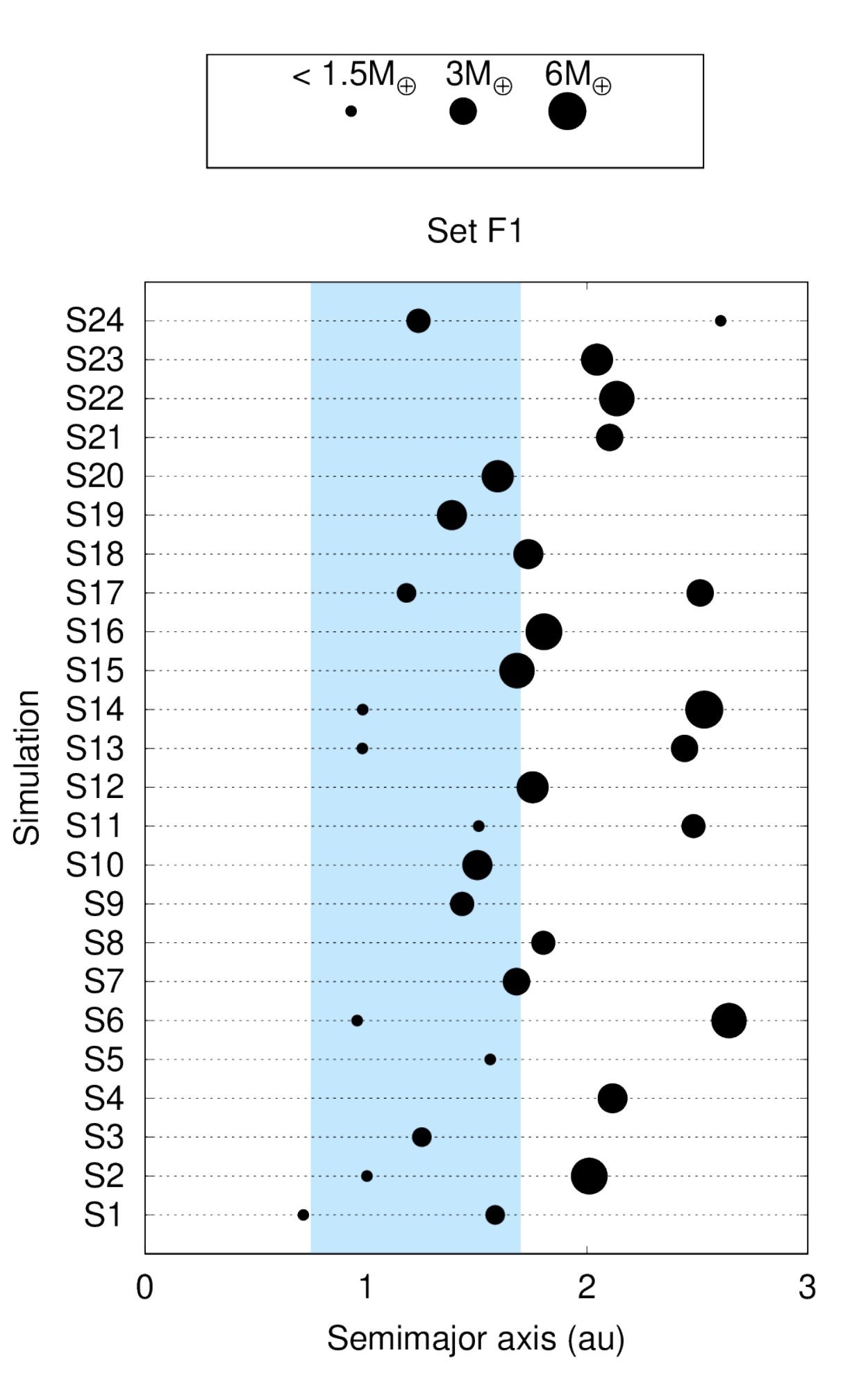}
 \includegraphics[angle=0, width= 0.33\textwidth]{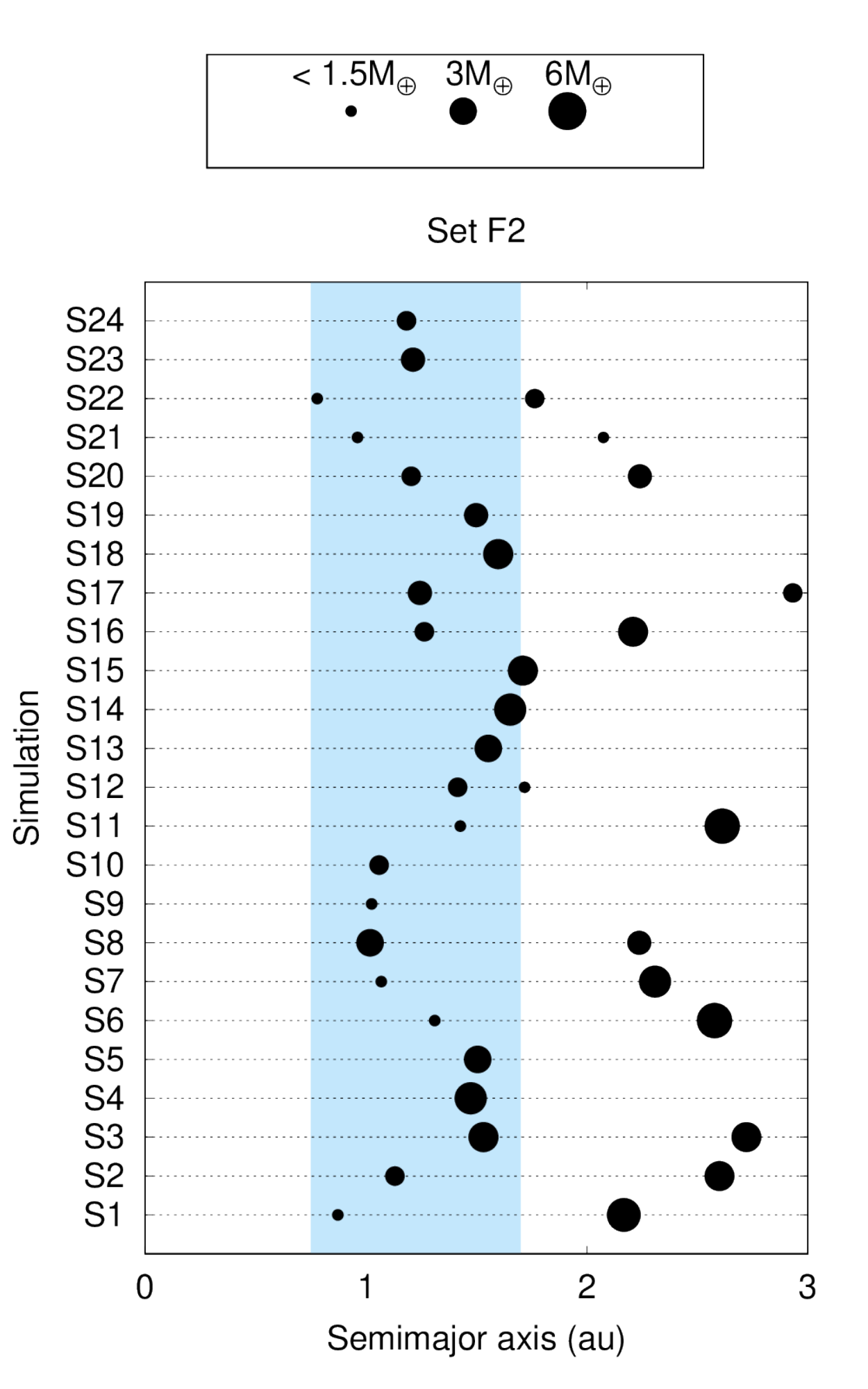}
 \includegraphics[angle=0, width= 0.33\textwidth]{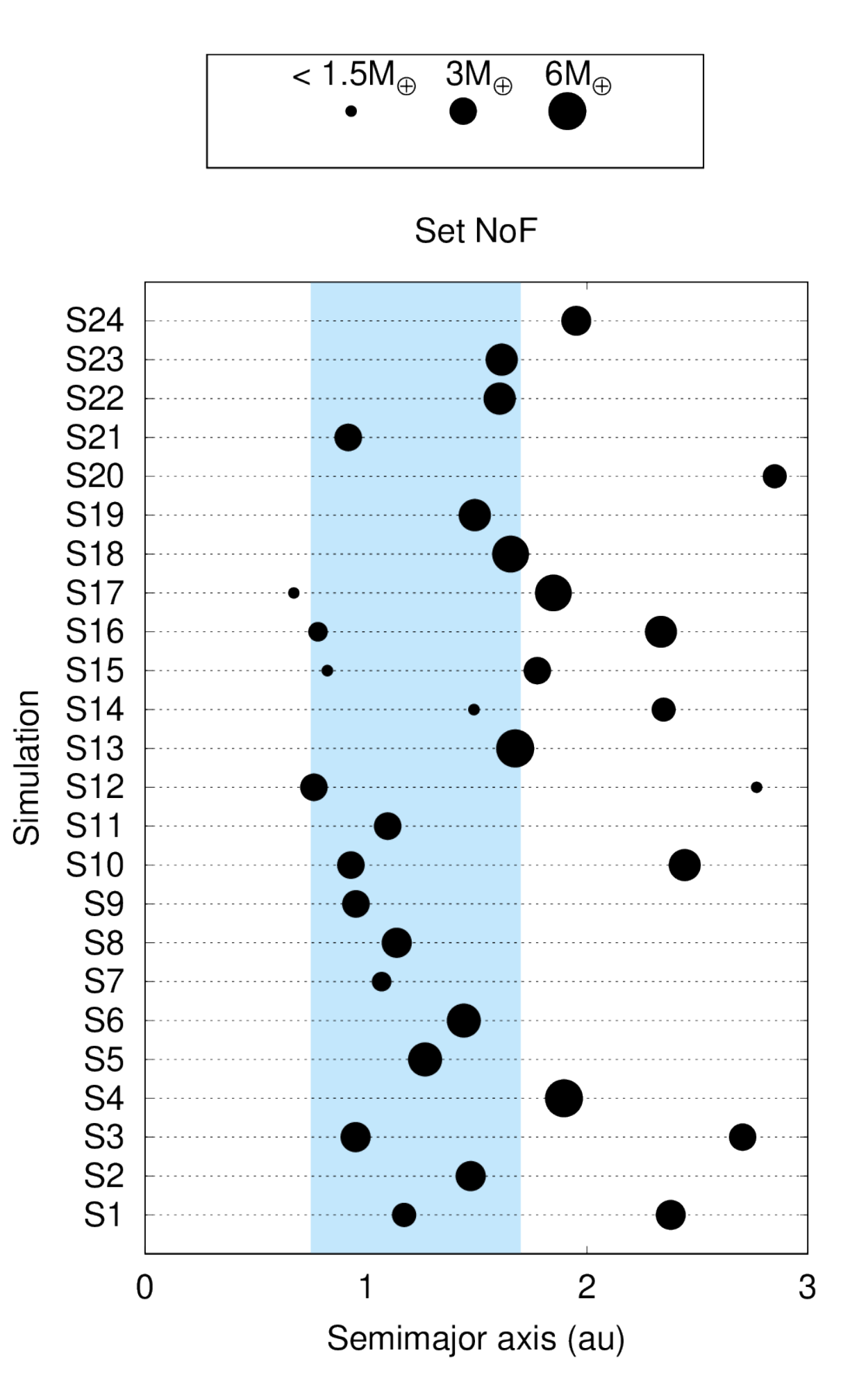}
 \caption{Planetary systems produced in the F1 (left panel), F2 (middle panel), and NoF (right panel) N-body simulations. The black circles represent the planets formed in the different numerical experiments, while each circle's size scales with the planet's mass, according to that indicated at the top of each panel. The sky-blue area illustrated in the different panels represents the HZ derived by \citet{Kopparapu2013a}.
}
\label{fig:arq}
\end{figure*}

The number of fragments per simulation generated in F1 runs ranged between 15 and 77. From the total number of fragments generated, erosive collisions and partial accretions produced from 1 to 77 fragments, while super-catastrophic collisions yielded between 1 and 50 fragments. For F2 simulations, where the fragments generation is more efficient, the number of fragments generated per simulation ranged between 148 and 791. In this case, erosive collisions and partial accretions produced from 1 to 242 fragments, while super-catastrophic collisions produced from 30 to 231 fragments. It is important to remark that the amount of fragments generated in a collision strongly depends on the total mass involved in it. Thus, it is possible to have more fragments generated from a partial accretion between massive bodies than from a super-catastrophic collision between small bodies. Moreover, several collisions involving larger bodies ended up in partial accretion collisions, leading to the generation of a considerably large number of fragments.

The bottom panel of Fig.~\ref{fig:nvst_ST} displays the temporal evolution of the number of embryos, without taking into account the fragments, for a representative simulation on each set. In this panel, it is possible to see that the evolution was slower in runs with fragmentation with respect to that without fragmentation.

Figure~\ref{fig:arq} illustrates the resulting planetary systems from F1 (left panel), F2 (middle panel) and NoF (right panel) simulations. Each formed planet is represented by a black filled circle, whose size is scaled with its mass. 

As a consequence of fragmentation, in sets F1 and F2, the formed planets were less massive than those obtained in the set NoF. Moreover, the final masses of the planets in the set F1 ranged between 0.18 M$_{\oplus}$ and 6.25 M$_{\oplus}$ while, for the set F2, these values ranged between 0.40 M$_{\oplus}$ and 5.50 M$_{\oplus}$. Lastly, for the set NoF, the final masses oscillated between  1.17 M$_{\oplus}$ and 6.74 M$_{\oplus}$.
This result was more evident around the HZ of the system. These observed discrepancies in the different regions of the system lead us to study in detail their physical and orbital properties in numerical simulations with and without fragmentation. 

\begin{figure}[ht!]
\centering
\includegraphics[angle=0, width= 0.49\textwidth]{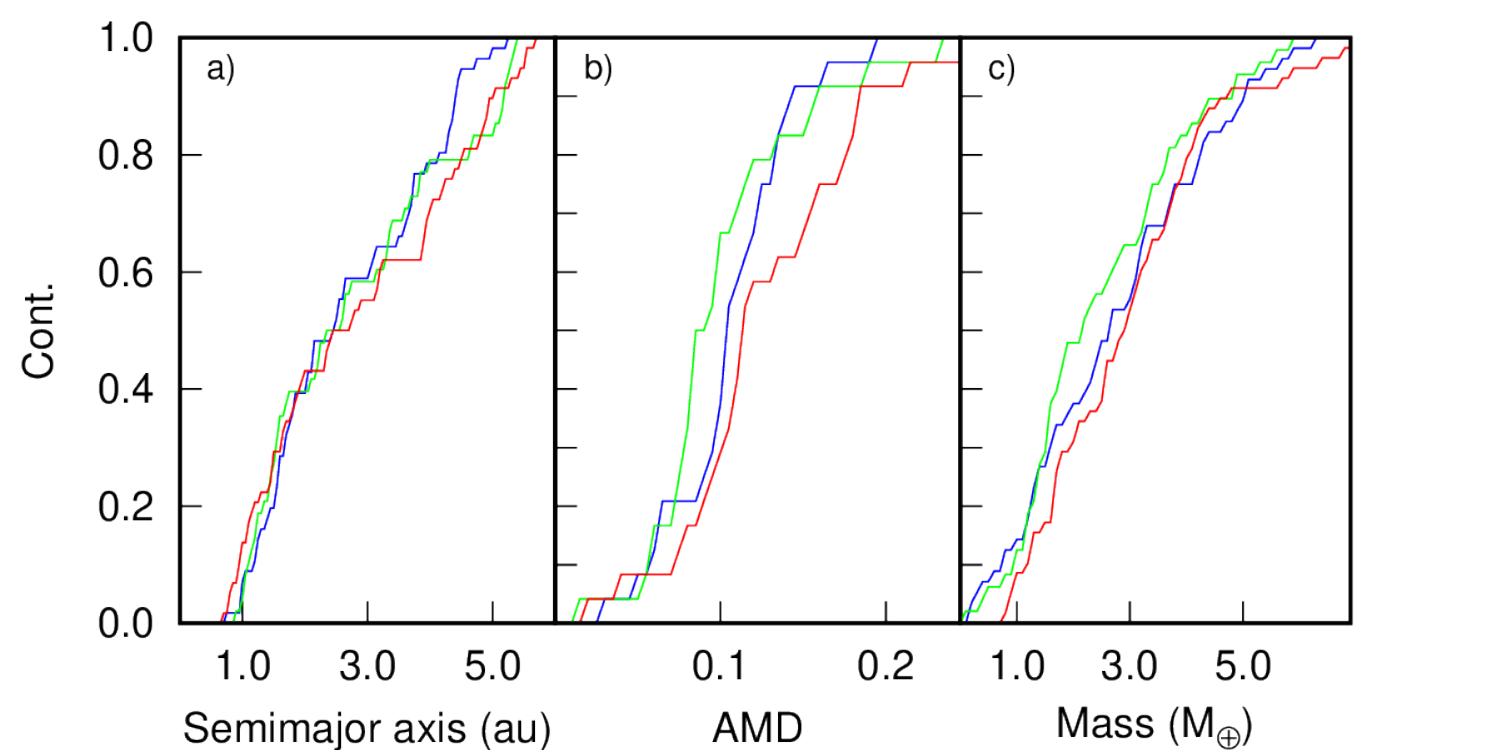}
\caption{Cumulative distribution of semimajor axis (left panel), angular momentum deficit (middle panel), and mass (right panel) for all the planets that survive in the simulations of sets F1 (blue), F2 (green), and NoF (red).}
\label{fig:multiplot}
\end{figure}

Figure \ref{fig:multiplot} illustrates the cumulative distribution of semimajor axis (left panel), angular momentum deficit (AMD) (middle panel), and mass (right panel) of all planets surviving in the simulations associated with sets F1 (blue), F2 (green), and NoF (red). The curves represented in the three panels show similar trends, but we can observe certain distinctive features with a more detailed exploration. The left panel shows that the fraction of surviving planets with a semimajor axis $a \lesssim$ 3 au is comparable in the three sets. However, the fraction of planets with a semimajor axis $a \lesssim$ 1 au was larger in simulations without fragmentation. In fact, eight planets survived in the set NoF, while 4 (3) planets formed in the set F1 (F2). We can observe these results with more detail in Fig. \ref{fig:arq}. 

As for the middle panel, we calculated the cumulative distribution of the normalized angular momentum (AMD) deficit for all the simulations performed. The AMD is a quantitative measure of an orbit's deviation of a given system from a co-planar, circular configuration (Laskar, 1997). We calculated the AMD given by Eq. \eqref{eq:AMD}.

\begin{equation}\label{eq:AMD}
    AMD = \frac{\Sigma_i m_i \sqrt{a_i}[ 1 - \sqrt{(1 - e_i^2)}\cos{i_i}]}{\Sigma_i m_i \sqrt{a_i}},
\end{equation}

\noindent{where} $m_i$, $a_i$, $e_i$, and $i_i$ are the mass, semimajor axis, eccentricity and inclination of the ith-body in the simulation, respectively.

We can observe that the fraction of simulations that included fragmentation exhibit smaller values of AMD than those without fragmentation. In fact, the maximum reached value of AMD for F1-F2 (NoF) simulations was of 0.16 (0.31). For AMD values lower than 0.05, the reader has to be aware that those values are not statistically meaningful, given that the represented values correspond to less than $5\%$ of the systems. The AMD values for the planets in these simulations are larger than those for the Solar System. This difference has to do with the fact that we have more massive bodies and more eccentric orbits.

In simulations with fragmentation, we observed that the eccentricity values were slightly lower than those in simulations without fragmentation. The overall AMD decrease reflects this behavior. As pointed out in \citet{Chambers2013}, the generation of collisional fragments increases the number of bodies in the system. This increase could lead to mild dynamical friction over the more massive bodies, reducing their eccentricity. This effect could be analogous to that observed by \citet{Obrien2006}, where they studied terrestrial planets formation under strong dynamical friction. In their work, they assumed a scenario with a classical model of accretion and a large number of planetesimals ($\sim$ 1000). The surviving planetesimals produce a decrease in the excitation levels of the formed planets due to dynamical friction. The effect they observed was larger than those in our work. This might be due to the difference in the number of small bodies involved. Moreover, our F2 scenarios showed a lower excitation with respect to F1 simulations. Since the number of fragments generated in F2 is considerably larger than in F1, the dynamical friction experimented by the final planets in the former was stronger than in the latter.

The mass distribution of the planets also offers interesting results. The right panel of Fig.~\ref{fig:multiplot} shows that the simulations with fragmentation formed planets with smaller masses than those obtained in runs without fragmentation.  

These results are consistent with those derived by \citet{Chambers2013} and \citet{Dugaro2019}, who showed that the final planets have somewhat smaller masses and eccentricities when a more realistic treatment of the collisional processes was included in the simulations. However, it is important to mention that the investigations carried out by \citet{Chambers2013} and \citet{Dugaro2019} took into account the effects of giant planets in the evolution of a system. According to this, we showed that the conclusions derived by \citet{Chambers2013} and \citet{Dugaro2019}, concerning the mass and eccentricity distributions of planets formed in simulations with and without fragmentation, seems to be insensitive to the presence of giant planets.

\begin{figure}[ht!]
\centering
\includegraphics[angle=0, width= 0.5\textwidth]{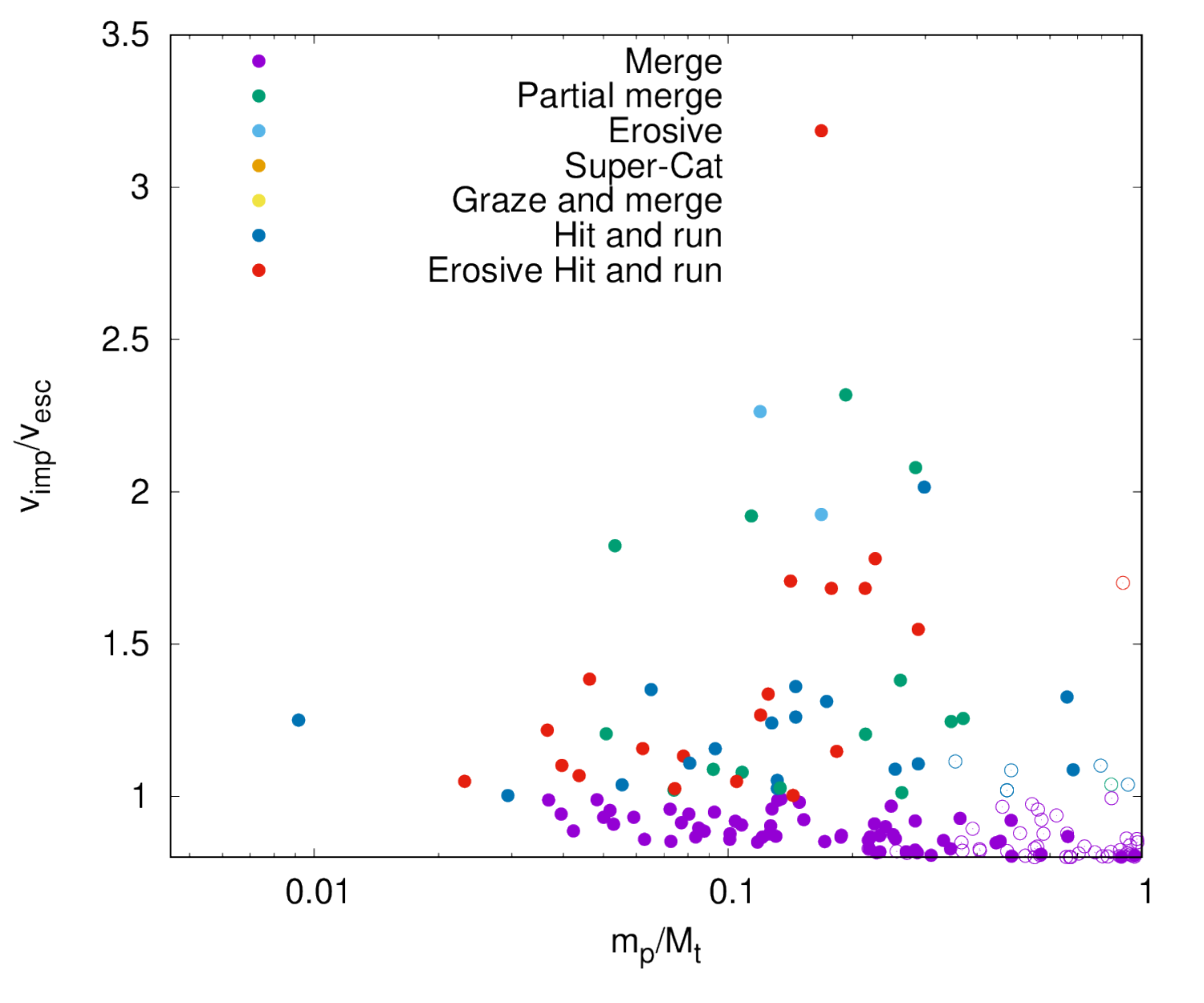}
\caption{Impact velocity, scaled to the mutual escape velocity, as a function of the projectile-to-target mass ratio for F1 simulations. Filled (hollow) symbols indicate if the impactor begun the simulation in the region interior (exterior) to the snow line. The color code indicates the different collision type. Here, only impacts on Earth analogs were considered.}
\label{fig:quintana}
\end{figure}

Figure \ref{fig:quintana} shows the impact velocity, scaled to the mutual escape velocity $v_{\text{imp}}/v_{\text{esc}}$, as a function of the projectile-to-target mass ratio $m_{\text{p}}/M_{\text{t}}$ for the set F1. The color code indicates the different collision types produced. This ratio corresponds to giant impacts onto Earth-analog planets, which are defined as those that have a final mass greater than 0.5 M$_{\oplus}$ and a final semimajor axis between 0.75 au and 1.5 au \citep{Quintana2016}. Additionally, we distinguish impactors that begun the simulation in the region interior (exterior) to the snow line represented by filled (hollow) symbols. The simulations for F1 and F2 scenarios presented similar behavior.

Our study shows that most of the collisions were produced at impact velocities $v_{\text{imp}}$ $\lesssim$ 2.0 $v_{\text{esc}}$, for values of $m_{\text{p}}/M_{\text{t}}$ between 0.009 and 1. We can split this mass range into two groups. In the first group, collisions at such velocities for values of $m_{\text{p}}/M_{\text{t}}$ between 0.05 and 1 are consistent with those obtained by \citet{Quintana2016}, who showed that impact events that involve bodies of comparable masses, had impact velocities $v_{\text{imp}}$ that were 1–2 times the mutual escape velocity $v_{\text{esc}}$. 
In this group, we can also observe that partial mergers, erosive collisions and hit-and-run encounters were more likely to have high impact velocities (above $\sim 1.8 v_{\text{imp}}/v_{\text{esc}}$) for $0.1 < m_{\text{p}}/M_{\text{t}} < 0.35$. Those impactors that originally formed in the region exterior to the snow line were more likely to produce perfect mergers and erosive collisions than other type of collisions as the reader can observe in the bottom-right corner of Fig. \ref{fig:quintana}.

The second group corresponds to collisions at $v_{\text{imp}}$ $\lesssim$ 2.0 $v_{\text{esc}}$ for values of $m_{\text{p}}/M_{\text{t}}$ between 0.009 and 0.05. Collisions in this mass-ratio range were not observed in the numerical simulations carried out by \citet{Quintana2016}. In fact, such authors suggested that smaller impactors onto more massive embryos reached impact velocities $v_{\text{imp}}$ that were 2-5 times the mutual escape velocity $v_{\text{esc}}$. We considered that this discrepancy was due to the differences between the scenario of study proposed in our research and that associated with \citet{Quintana2016}'s work. In fact, while we analyzed the formation and evolution of terrestrial-like planets in absence of giant planets around sun-like stars, the systems studied by \citet{Quintana2016} included the effects of outer planets analogs to Jupiter and Saturn. Finally, Fig.~\ref{fig:quintana} shows that there is not a clear distinction between F1 and F2 simulations concerning  $v_{\text{imp}}/v_{\text{esc}}$ for all the projectile-to-target mass ratio $m_{\text{p}}/M_{\text{t}}$.

\begin{figure}[ht!]
\centering
\includegraphics[angle=0, width= 0.49\textwidth]{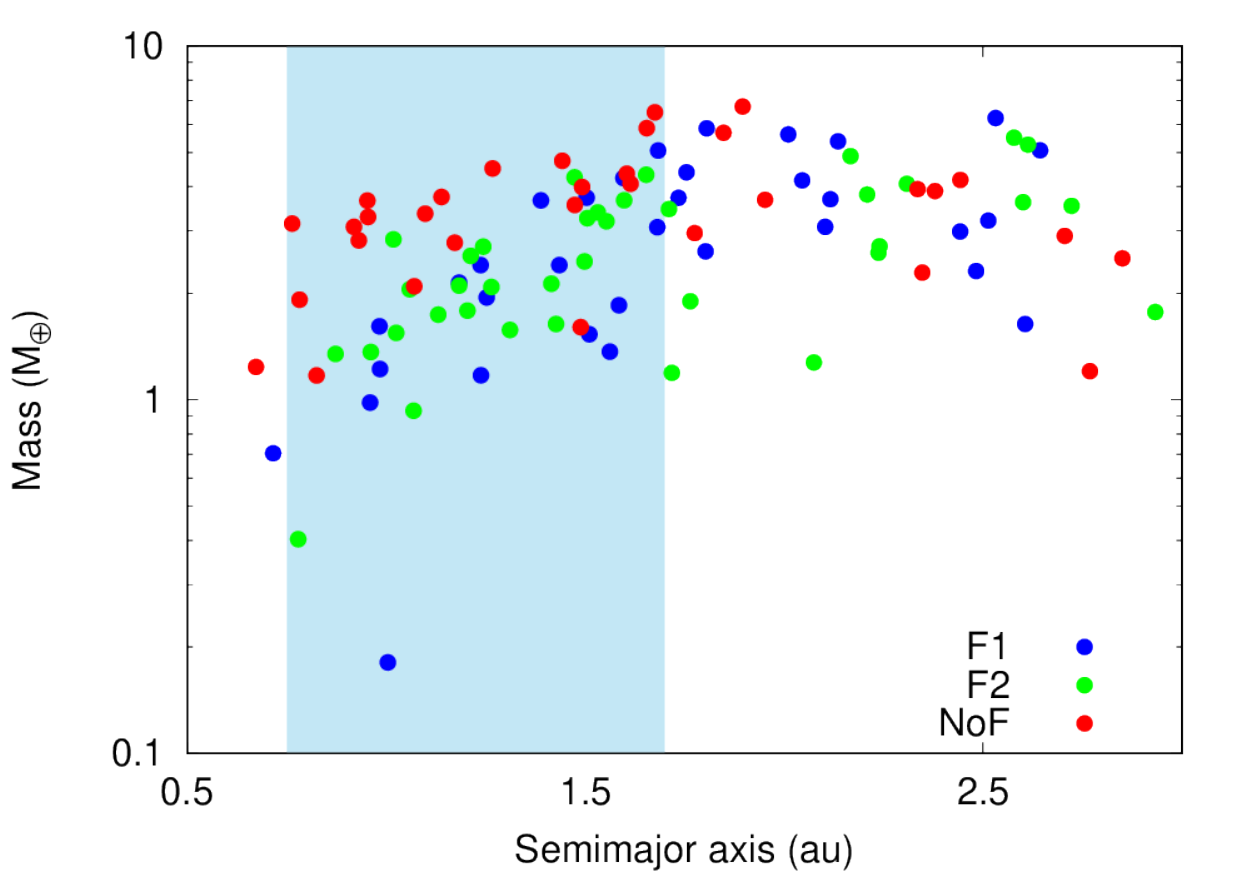}
\caption{Final mass of the planets that survived between 0.5 au and 3 au, after 400 Myr, in F1, F2, and NoF simulations. Planets formed in F1 (F2) N-body experiments are represented by blue (green) filled circles, while those produced in NoF simulations are illustrated by red filled circles. The sky-blue shaded area represents the HZ for a solar-type star.}
\label{fig:distr_masa_final}
\end{figure}

Finally, we analyzed the properties of the planets produced in the different sets of N-body simulations of our research. Fig.~\ref{fig:distr_masa_final} shows the final mass of the planets formed in F1 (blue circles), F2 (green circles), and NoF (red circles) simulations as a function of the final semimajor axis. As we have already mentioned in our analysis of Fig.~\ref{fig:multiplot}, the mass of the resulting planets in NoF simulations were somewhat greater than those obtained in F1 and F2 simulations. This is a natural result since there was no mass loss in the collisions of the NoF simulations, all of which were treated as perfect mergers. The most massive planet resulting from our numerical experiments had a mass of 6.6 M$_{\oplus}$ and was produced from a NoF simulation. On the contrary, the less massive planets in our study were formed from N-body experiments that include fragmentation. In fact, F1 (F2) simulations produced a planet with a final mass of 0.18 M$_{\oplus}$ (0.4 M$_{\oplus}$). A distinctive feature observed in Fig.~\ref{fig:distr_masa_final} is related to the existence of a large number of planets with masses larger than 1 M$_{\oplus}$ in all our N-body experiments, regardless of the collisional treatment. It is important to remark that such planets, which were produced in absence of giants, were significantly more massive than those obtained in \citet{Dugaro2019}, who included outer planets analogs to Jupiter and Saturn in their N-body experiments. 
The sky-blue shaded area of Fig.~\ref{fig:distr_masa_final} represents the HZ adopted in the present research. According to this, a large number of planets survived in the HZ of the system in F1, F2, and NoF simulations. Such planets are of significant interest due to their potential habitability, for which we carry out in the next section a detailed analysis concerning their physical and orbital properties.

\subsection{Planets surviving in the habitable zone}

In the present section, we studied the formation, evolution, and physical properties of the planets that survived in the HZ of the system in F1, F2, and NoF N-body simulations.

First of all, it is imperative to analyze the formation region of a planet that survived in the HZ to understand its final physical properties. We could distinguish two different types of planets surviving in the HZ in the F1, F2, and NoF simulations. Such a classification was based on the initial location of those planets in the disk. Following \citet{Dugaro2019}, we refer to class A (class B) planets as those in the HZ whose accretion seed started the simulation inside (beyond) the snow line\footnote{For a detailed definition concerning the accretion seed of a planet in N-body simulations with and without fragmentation see \citet{Dugaro2019}.}. 

\begin{figure}[ht!]
\centering
\includegraphics[angle=0, width= 0.49\textwidth]{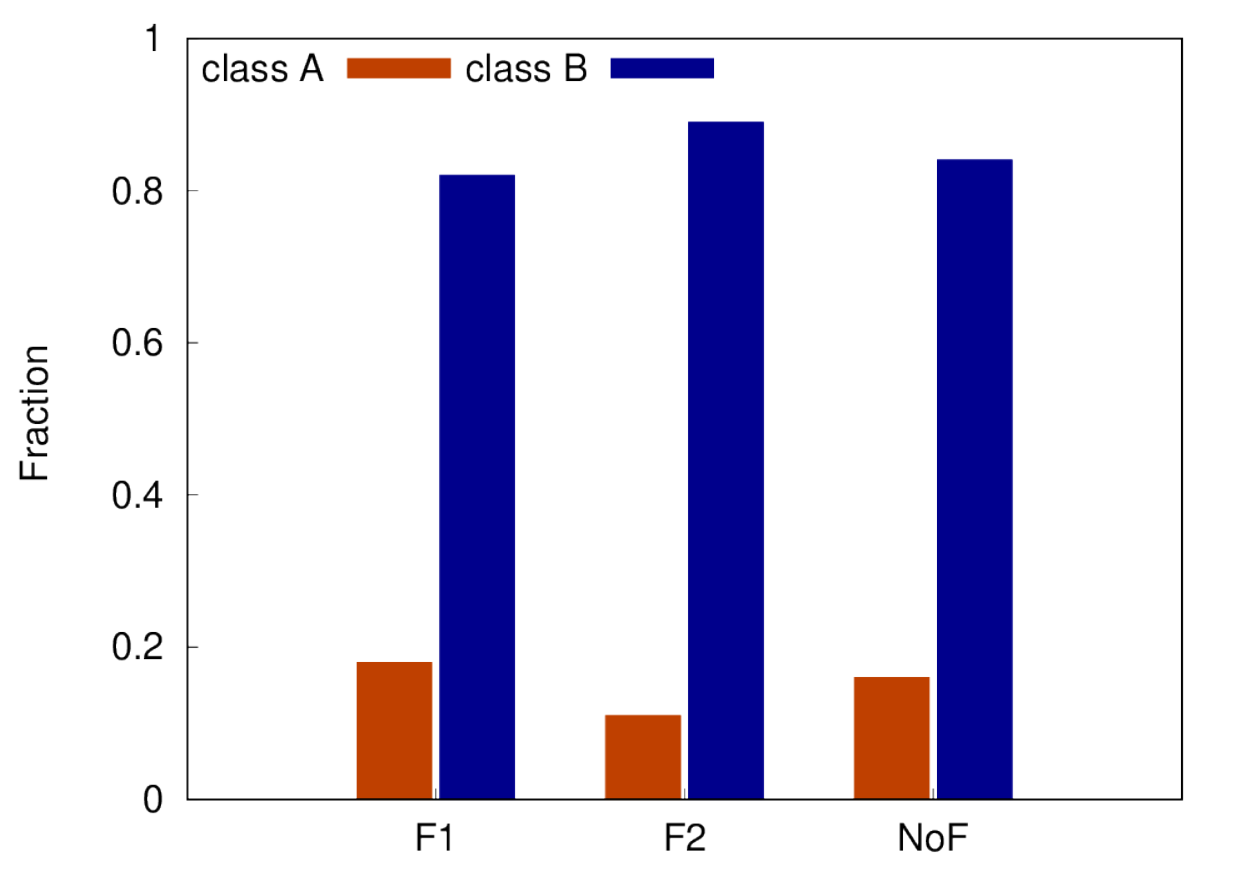}
\caption{Fraction of planets surviving in the HZ in F1, F2, and NoF simulations. Dark-orange (dark-blue) bars represent class A (class B) planets.}
\label{fig:histo_plan_zh}
\end{figure}

Figure~\ref{fig:histo_plan_zh} shows the number of class A and class B planets, normalized to the total number of planets surviving in the HZ, in the F1, F2, and NoF simulations. It can be seen that the fraction of planets formed is similar for the three scenarios. It is also clear that the fraction of class B planets is substantially larger than that of class A. Thus, we decided to focus our investigation on the physical and dynamical properties of the class B planets formed in the three different sets of numerical simulations.

Class B planets presented different evolutionary histories in simulations with and without fragmentation, for which it is interesting to analyze how such planets grew throughout their evolution. 
As a consequence of the implementation of a more realistic collision treatment in F1 and F2 simulations, not all collisions experienced by the planets resulting from those runs were perfect mergers.
However, most of the collisions experienced by class B planets were perfect mergers and partial accretions, and few of them were erosive impacts, leading to a mass loss throughout their entire evolution. Our results suggest that the mass evolution over time of the planets seems not to depend on the adopted value of $M_{\text{min}}$. This result is consistent with those obtained by \citet{Wallace2017} who studied, through N-body simulations, rocky planet formation for different values of $M_{\text{min}}$ at small semimajor axes.
The class B planets that resulted from F1 (F2) simulations had final masses ranging from 0.98 (0.93) M$_{\oplus}$ to 5.05 (4.30) M$_{\oplus}$, while the masses of class B planets produced in NoF simulations ranged between 1.60 M$_{\oplus}$ and 6.5 M$_{\oplus}$. It is important to remark that the contribution of the collisional fragments to the final mass of class B planets in F1 and F2 simulations was not significant.

\begin{figure}[ht!]
\centering
\includegraphics[width= 0.45\textwidth,angle=0]{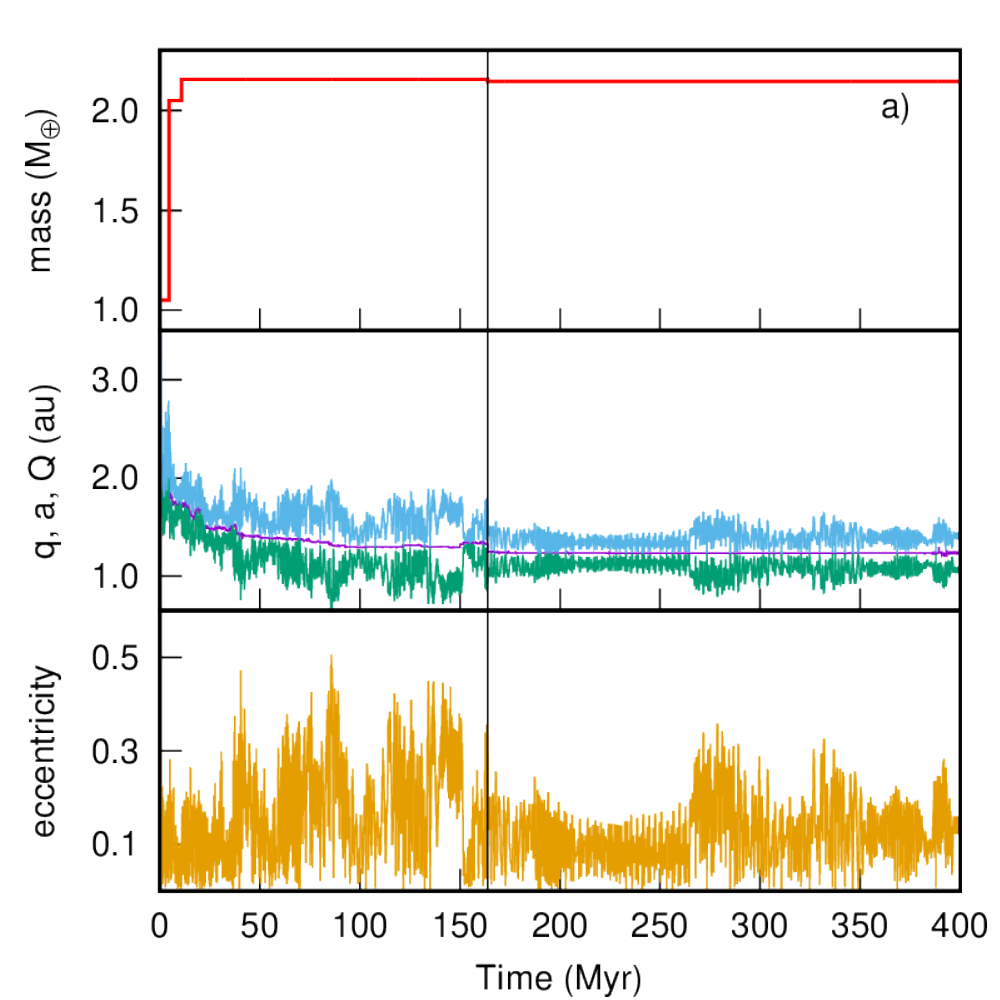}
\includegraphics[width=0.45\textwidth,angle=0]{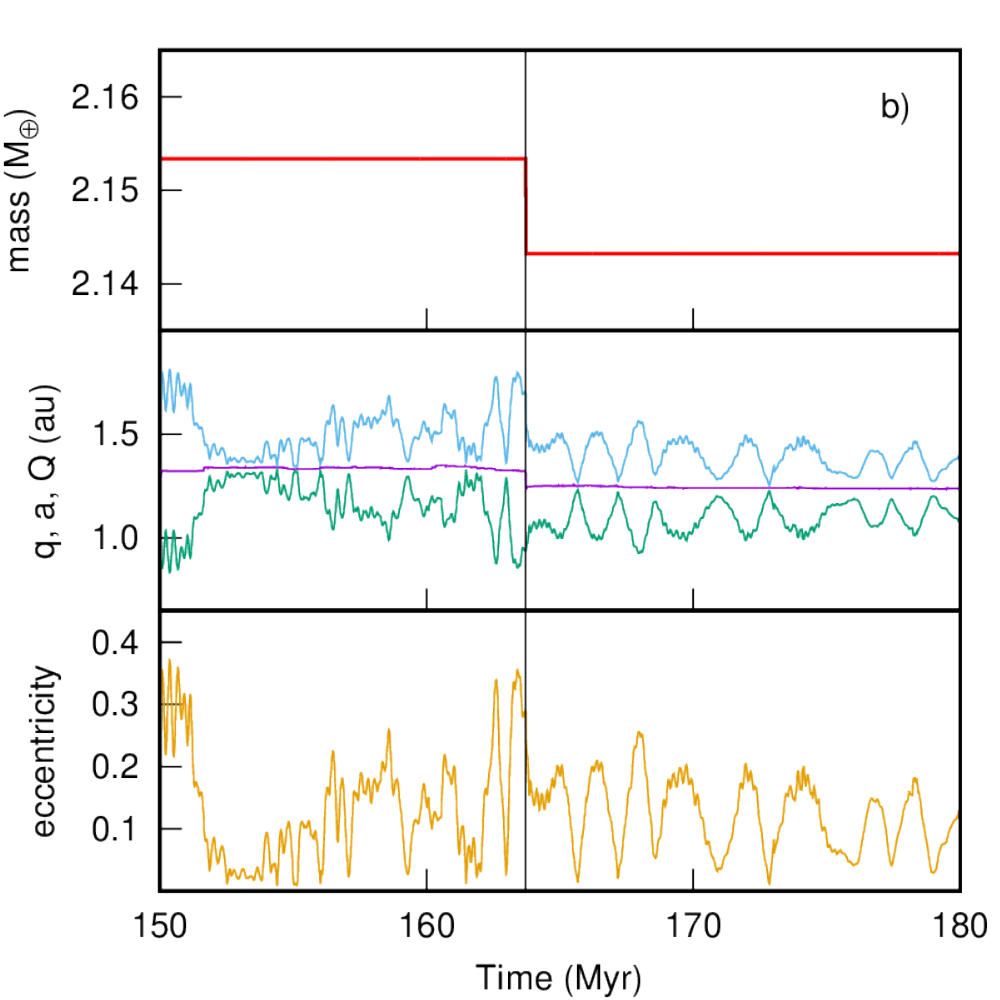}
\caption{Physical and orbital evolution in time of a class B planet from set F1. Panel a): the top panel corresponds to the mass' evolution. In the middle panel, we display the evolution of pericentric distance (q), semimajor axis (a), and apocentric distance (Q) with light blue, purple, and green traces, respectively. The bottom panel shows the evolution of eccentricity. Panel b) displays a temporal magnification around the time corresponding to an erosive collision with a Mars-mass projectile. The color code is analogous to Panel a). The black vertical line in both panels indicates when 
the planetary embryo suffered an erosive collision.}
\label{fig:growth}
\end{figure}

An example of growth history and dynamical evolution of a class B planet can be observed in Fig. \ref{fig:growth}. The top panel of Fig. \ref{fig:growth} a) shows the temporal evolution of mass, while pericentric distance, semimajor axis, and apocentric distance are plotted in the middle panel. Lastly, the eccentricity evolution over time is in the bottom panel. The planetary embryo started the simulation with a semimajor axis of 3.1 au and a mass of 1.05 M$_{\oplus}$. Throughout its dynamical evolution, the body suffered a series of impacts that changed its mass and orbital parameters. In the first 20 Myr, the embryo had 2 perfect mergers that increased its mass by a factor of two. The body suffered an erosive collision around 163 Myr with a Mars-mass projectile that caused a minor effect in its mass, as observed in the top panel of Fig. \ref{fig:growth} a). The planet's orbital evolution suffered several changes as well. In the first 50 Myr of evolution, the object's semimajor axis decreased from 3.1 au to 1.38 au due to the impacts previously mentioned, close encounters, and a hit-and-run collision. Lastly, an erosive collision reduced its semimajor axis to 1.23 au, preserving that value for the rest of the simulation. The three panels of Fig. \ref{fig:growth} b) represent the physical and dynamical evolution for the same object, displayed in a more suited timescale to better appreciate its mass and orbital parameters changes due to the erosive collision.

\begin{figure}[ht!]
\centering
\includegraphics[width= 0.45\textwidth,angle=0]{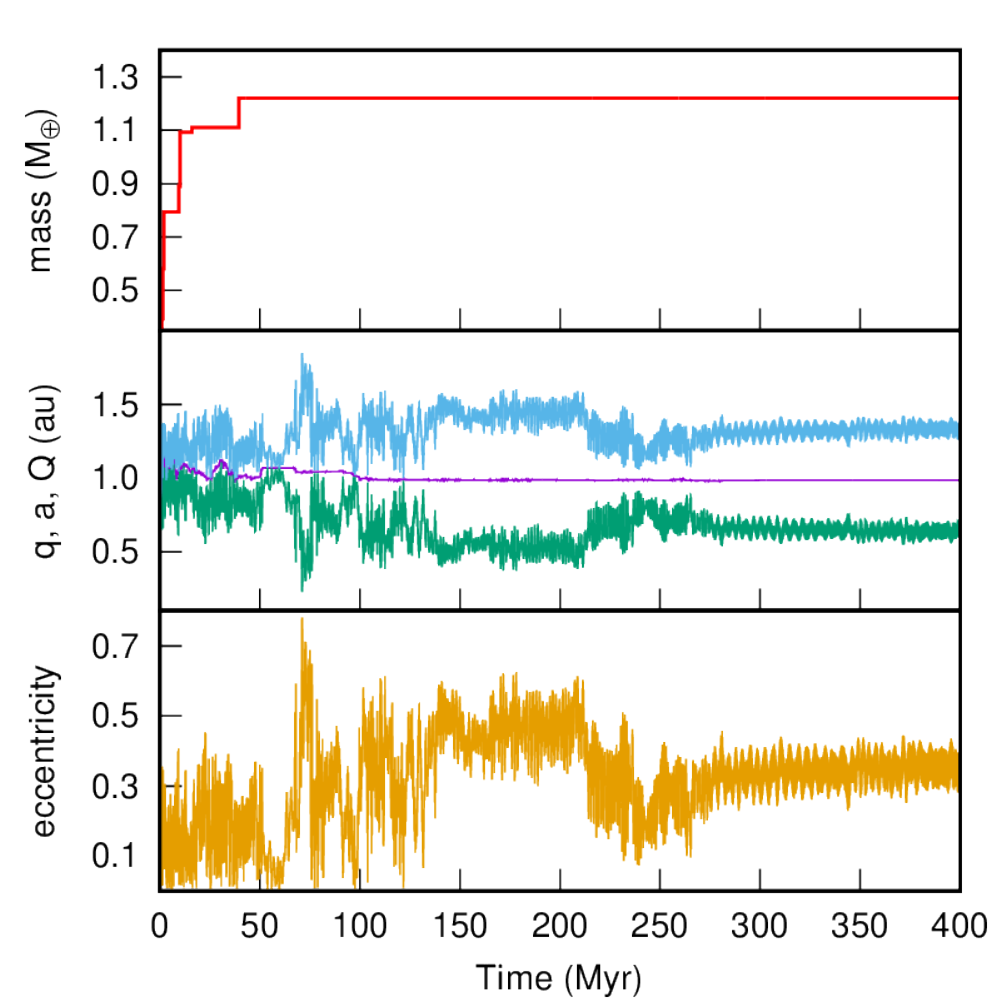}
\caption{Evolution in time of the mass, pericentric distance (q), semimajor axis (a), and apocentric distance (Q) and eccentricity in the panels top, middle and bottom, respectively, for a class A Earth analog.}
\label{fig:growth2}
\end{figure}

Although we are interested in class B planets, a similar analysis can be done to a class A Earth analog. We show, in Fig. \ref{fig:growth2}, the physical and dynamical evolution for a planet that grew in the region interior to the snow line. The body started with a mass of 0.07 M$_{\oplus}$ and ended with 1.2 M$_{\oplus}$ as a consequence of 10 perfect mergers and 2 partial accretions. Additionally, the body suffered 4 hit-and-run collisions. As the reader can observe in the middle panel of this figure, the object's semimajor axis didn't change significantly over time.

The different growing histories of these planets give the idea of the importance of radial mixing when studying the final physical properties of such objects. Within the physical properties of the formed planets, we are interested in studying the gain/loss balance of final water content. The following section addresses this topic in more detail for all the class B planets in F1, F2 and NoF simulations.

\subsection{Water content of the HZ planets}

Life as we know has been possible thanks to the water (among other compounds) present in the Earth. Analysis of the water content of the planets that survive in the HZ of a given system is a captivating topic to study since it allows us to understand their potential astrobiological interest \citep{Meech2019}. 

In the scenarios presented in this paper, we tracked each planet’s water mass fraction throughout the entire simulation and determined how it changed over time due to the successive collisions that it experienced. 

In runs that treated all collisions as perfect mergers, the total mass and water contents of the interacting body system were assigned to the resulting body from the perfect accretion outcome. For simulations with fragmentation, we developed a prescription for distributing both the projectile and target water content between the largest remnant and the fragments generated. To do this, we adopted the procedure presented in \citet{Dugaro2019}, who took into account the mantle stripping models proposed by \citet{Marcus2010}, which are described as follows:

\begin{itemize}
\item Model 1: It assumes that all the escaping mass is the lightest material (in this case, water), first, from the projectile and then from the target. Then, rocky material from such bodies escapes in the same order. 
\item Model 2: It assumes that the mass that escapes is water (first) and rocky material (second) from the projectile. Then, water and rocky material from the target escape in the same order. 
\end{itemize}

From Fig.~\ref{fig:histo_plan_zh}, we see that more than 80\% of the planets that survived in the HZ of the system in each of the sets F1, F2, and NoF are those defined as class B. Given that these types of planets started the numerical simulations beyond the snow line, they have very high initial fractions of water by mass. Their final water fractions can be significantly large. These fractions depend on their initial water fractions and their feeding zones. A very important point of our research is to determine the final water contents acquired by class B planets formed in simulations with and without fragmentation, to understand the sensitivity of their physical properties to the collisional model adopted.

\begin{figure*}[ht!]
\centering
\includegraphics[angle=0, width=1.0\textwidth]{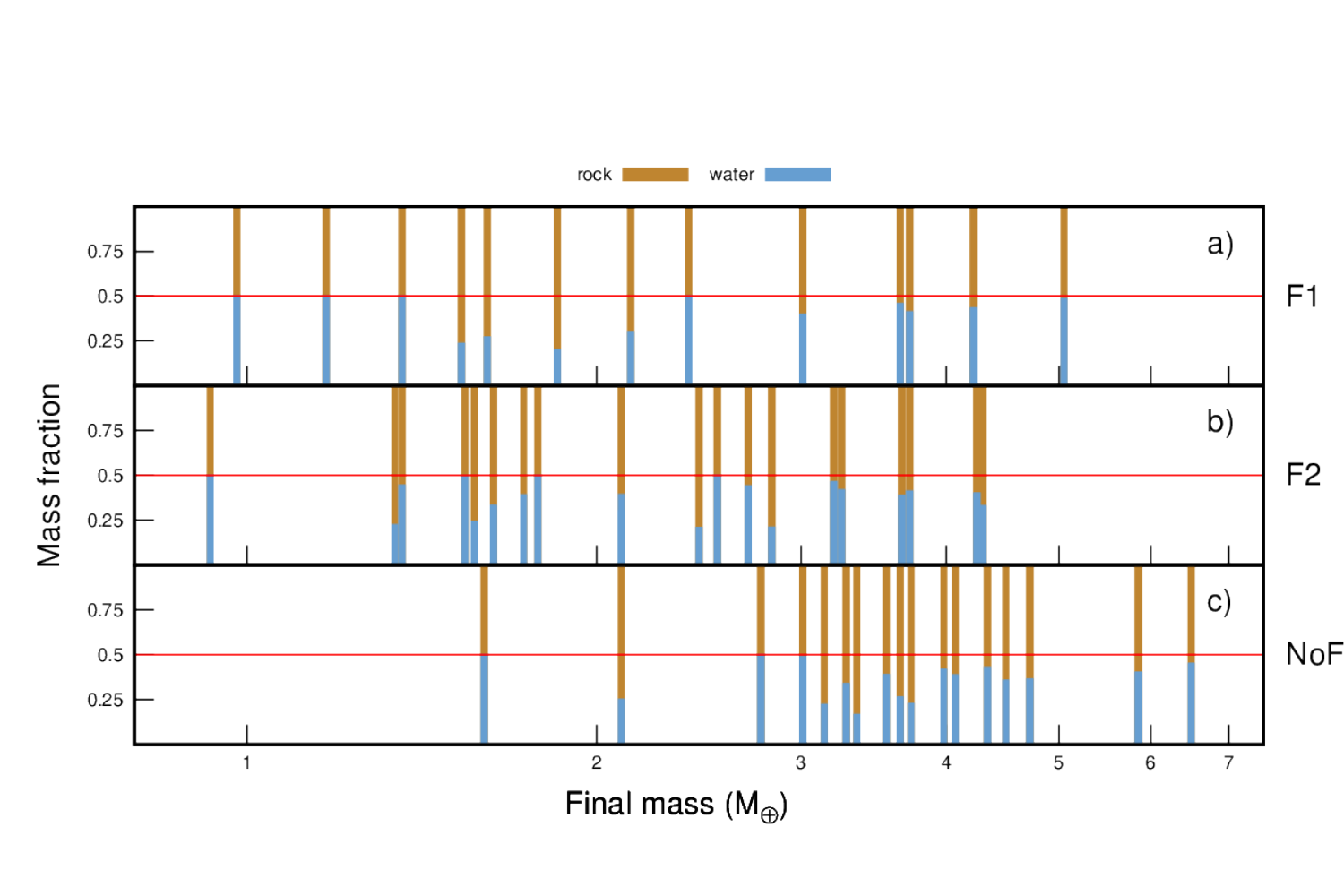} 
\caption{Fraction of rock/water of the class B planets that survived in the HZ after 400 Myr of evolution for F1 (top), F2 (middle), and NoF (bottom) simulations. Dark-yellow (dark-blue) bars indicate the fraction of rocky (water) content present in each planet. The red line indicates the initial water fraction at the beginning of the simulation.} 
\label{fig:agua_final}
\end{figure*}

Figure~\ref{fig:agua_final} illustrates the fractions of rock (dark-yellow bars) and water (dark-blue bars) of the class B planets formed in the three different sets of numerical simulations F1 (top panel), F2 (middle panel), and NoF (bottom panel). The horizontal red dashed line represents the initial fraction of water of all class B planets of our simulations.

In particular, the bottom panel of Fig.~\ref{fig:agua_final} indicates that the class B planets formed in the NoF simulations are true water worlds. In fact, such planets show final fractions of water by mass ranging from 0.16 to 0.5, with a median value of 0.38. This result is consistent with previous studies based on N-body simulations without fragmentation, which analyzed the physical and orbital properties of terrestrial-like planets in the HZ in absence of gaseous giants \citep[e.g.][]{deElia2013, Ronco2014, Dugaro2016, Zain2018}. 

In the present investigation, we wanted to determine if the inclusion of a more realistic collisional treatment in the N-body simulations may represent a barrier to the formation of water worlds in the HZ. To do this, we initially adopted the previously defined Model 1 to track the evolution of water in each impact event. From this, the top and middle panels of Fig.~\ref{fig:agua_final} illustrate the final fractions of rock and water of the class B planets formed in the F1 and F2 simulations, respectively, using Model 1. According to this, the final fractions of water by mass of the class B planets produced in the F1 (F2) simulations ranged from 0.2 (0.21) to 0.5 (0.5), with a median value of 0.44 (0.4). Then, we repeated this last process using Model 2 to track the evolution of water in each collision and recalculated the final fraction of water by mass of the class B planets formed in both sets. We did not find significant differences in the water content of those planets. Thus, our results show that the high final fraction of water by mass of class B planets produced in simulations with fragmentation is determined by the primordial water content, and it does not strongly depend on the water loss models used in the present study.   

Figure~\ref{fig:agua_final} also allows us to observe an interesting result concerning the final mass and water content of the class B planets formed in our simulations. The top and middle panels show that simulations that included a more realistic collisional treatment formed more class B planets with masses $\lesssim$ 2.0 M$_{\oplus}$ than those simulations that only assumed perfect mergers. From this, our results indicate that the fragmentation played a key role in the formation of Earth-analog planets with very high final water contents in the scenarios of study of the present research. 

Table \ref{tab:resumen} shows a summary of the physical properties of the class B planets formed in F1, F2, and NoF simulations using Model 1.  
\begin{table*}[ht!]
  \begin{center}
    \begin{tabular}{c|c|c|c|c|c|c} 
    \hline
      \textbf{Name} & \textbf{N$^{\circ}$ of HZ planets} & \textbf{N$^{\circ}$ of Class B planets} & \textbf{Masses (\textbf{Class B})} & \textbf{Mass median}           & \textbf{Wt. fr. (Class B)} & \textbf{Wt. fr. median}\\
      \hline
               F1   & 17                    & 14               & 0.98 - 5.05 [M$_{\oplus}$]         &  2.27 [M$_{\oplus}$]  & 0.20 - 0.50               &  0.44 \\
               F2   & 23                    & 20              & 0.93 - 4.30 [M$_{\oplus}$]         &  2.28 [M$_{\oplus}$]  & 0.21 - 0.50               &  0.40 \\
               NoF  & 20                    & 17               & 1.60 - 6.5 [M$_{\oplus}$]         &  3.63 [M$_{\oplus}$]  & 0.16 - 0.50               &  0.38 \\
      
      \hline
    \end{tabular}
    \vspace{5mm}
       \caption{Total number of HZ planets (column 2), number of class B planets (column 3), mass range of class B planets (column 4) and its median value (column 5), range of values associated with the fraction of water by mass of class B planets (column 6) and its median value (column 7). The water contents computed in the present table were obtained using Model 1.}
       \label{tab:resumen}
  \end{center}
\end{table*}

\section{Limitations of the model}
\label{sec:limitations}

It is important to mention that our numerical model has several limitations, which must be discussed. First of all, the zero time of our N-body simulations corresponds to the instant at which the gas disk has been fully dissipated from the system. Our initial conditions are very simple and they are based on classical models of embryo growth \citep{Kokubo1998,Kokubo2000,Kokubo2002}. In this sense, it is necessary to remark that we only consider planetary embryos in our initial populations, while the effects of a reservoir of planetesimals were not included.

More realistic numerical simulations should account for the effects of the gaseous component of the disk over the solid material. Detailed models developed during the last years can be incorporated in N-body simulations to simulate the evolution of the gas disk \citep{Bitsch2015,Ida2016,Liu2019}. According to this, initial conditions should include a population of planetesimals/embryos with masses consistent with the pebble accretion \citep{Johansen2010}, while the numerical code that models the evolution of the bodies in the gaseous phase should account for physical processes such as pebble accretion \citep{Ormel2010,Lambrechts2012}, gas accretion/erosion \citep{Ginzburg2016}, orbital migration, and damping of eccentricities and inclinations due to planet-disk interactions \citep{Ida2020}. We are aware that the incorporation of such physical mechanisms in the numerical model should lead us to derive more realistic initial conditions at the end of the gaseous phase.

We must also remark that the collisional model proposed in the present work has been applied to bodies with different densities and compositions. In our model, embryos interior (exterior) to the snow line are silicate (water)-rich bodies at the beginning of the simulations. Moreover, the water ice-to-rock ratio of each body may change over time due to the collisional processes between objects from different regions of the disk. In this sense, \citet{Leinhardt2012} suggested that the velocity of the largest remnant is likely to be sensitive to internal structure and composition of the bodies. Moreover, they indicated that the role of tidal effects during collisions or in close encounters may play an important role during the fragmentation of planetary bodies.
From these considerations, we consider that future works should be aimed at improving the collisional treatment of bodies with different physical properties in N-body simulations.

We would also like to mention that the model used to track the evolution in time of the water content of the bodies is very simple. Such a model assumes that, when a collision with fragments generation occurs, the first escaping mass is the lightest material, which is represented by water in our work \citep{Marcus2010}. The use of more realistic results of volatile transport and loss derived from hydrodynamics studies, such as those developed by \citet{Dvorak2015}, should lead to obtaining more precise abundances of water on the planets resulting from our simulations.

Finally, as \citet{Chambers2013} and \citet{Wallace2017} described in their works, our numerical algorithm also conserves the total mass of the interacting bodies after each collision. The fragmented mass is divided into the largest remnant and the generated fragments. Future work should incorporate a factor of mass removal, assuming that most fragments are ground to smaller sizes in a collisional cascade, and then removed by radiation forces before they can be accreted by planetary embryos \citep{Mustill2018}.

From the limitations described above, it is important to clarify two points of our research. First, we decide to adopt a simple model for the development of our study to focus on the differences in the physical and dynamical properties of planets formed from a numerical model based on perfect mergers and another one that includes a more realistic treatment of the collisional processes. Second, the limitations of our model indicate that the true diversity of physical and dynamical properties of the terrestrial-like planets formed in absence of giants around Sun-like stars will be much larger than we have derived in the present study.


\section{Discussion and conclusions}
\label{section:discusion}
In the present research, we carried out a study concerning the formation and evolution of terrestrial-like planets and water delivery in the HZ, in the absence of gaseous giants around solar-type stars, during the late-stage accretion phase of terrestrial planet formation. In particular, we developed a comparative analysis between N-body simulations that included a realistic collisional prescription and a classical model of accretion, which treated all collisions as perfect mergers. To do this, we made use of the \emph{D3} code developed by \citet{Dugaro2019}, which includes hit-and-run collisions and planetary fragmentation. This numerical integrator is based on the works carried out by \citet{Chambers1999}, \citet{Leinhardt2012}, \citet{Genda2012}, \citet{Chambers2013}, and \citet{Mustill2018}. 

In general terms, the final planetary systems produced in N-body simulations with and without fragmentation showed similar results concerning the number of planets, semimajor axis distribution, and growth rates. However, there were differences in the physical and orbital properties of the terrestrial-like planets formed in the two models. One of these differences is that the planets produced in simulations with fragmentation had final masses lower than those obtained in runs without fragmentation. The other difference is that the systems obtained when a more realistic collisional treatment was included presented lower excitation in comparison with those using the classical model of accretion.
These results show that the implementation of a more realistic collision treatment and a fragmentation regime yields a difference in the systems' dynamical evolution.

We are aware of the limitations of the proposed model as we described in Section \ref{sec:limitations}. However, this simplified model allowed us to obtain those differences in the results. This leads us to conclude that the implementations carried out in our code constitutes an important step in a new generation of N-body codes. Moreover, future works that address the dynamical processes involved in planetary formation should be carried out using this kind of models.\\

Our results suggest that the final systems produced in simulations that include fragmentation, where the minimum permitted mass for the fragments $M_\text{min}$ was varied, are broadly similar. Moreover, we obtained comparable results for the frequency of occurrence of the different types of collisions, the temporal evolution of the number of embryos, the number of resulting planets, the semimajor axis, eccentricity and mass distribution, the kind of planets formed in the HZ, and the final fraction of water by mass of the class B planets.  
However, the selection of $M_\text{min}$ may be important in the analysis of, for instance, the atmospheric mass loss or differentiation processes, among others, which have not been explored in the present research. Moreover, the value adopted for $M_\text{min}$ may be relevant in the final physical properties of class A planets of a given system, in which the collisional fragments play a key role in their evolutionary history \citep{Dugaro2019}.

The results obtained from our simulations suggest that fragmentation is not a barrier to water world formation in the HZ for the classic model of embryo accretion. This investigation may help us to strengthen our understanding of the physical properties of potentially habitable terrestrial-like planets in the Universe.

A refined study could be added in future works regarding the atmospheric loss since giant impacts produce a blow-off of atmospheric compounds in each collision. Given the stochastic nature of giant impacts on the final planets of a planetary system, it may lead to a wide variety of atmospheric removal fraction \citep{Stewart2014}.

A detailed treatment of collisions and water transport in terrestrial planet formation has great relevance to understanding the vast multiplicities and architectures of planetary systems in the Universe, and the different physical properties of new exoplanets discoveries. 

\begin{acknowledgements}
{This work was partially financed by Agencia Nacional de Promoci\'on Cient\'{\i}fica y Tecnol\'ogica (ANPCyT) by PICT 2016-2635 and PICT 201-0505, and by Universidad Nacional de La Plata (UNLP) through PID G144. Moreover, we acknowledge the financial support by Facultad de Ciencias Astron\'omicas y Geof\'{\i}sicas de la Universidad Nacional de La Plata (FCAGLP-UNLP) and Instituto de Astrof\'{\i}sica de La Plata (IALP) for extensive use of their computing facilities. The authors thank the anonymous reviewer for a helpful and insightful report which helped to improve this article.}
\end{acknowledgements}
\section*{ORCID-ID}
Agustin Dugaro: https://orcid.org/0000-0002-7188-0848
\bibliographystyle{aa} 
\bibliography{solo_terrestres.bib} 

\end{document}